\documentclass{article}

\usepackage{PRIMEarxiv}

\usepackage[utf8]{inputenc} % allow utf-8 input
\usepackage[T1]{fontenc}    % use 8-bit T1 fonts
\usepackage{hyperref}       % hyperlinks
\usepackage{url}            % simple URL typesetting
\usepackage{booktabs}       % professional-quality tables
\usepackage{amsfonts}       % blackboard math symbols
\usepackage{nicefrac}       % compact symbols for 1/2, etc.
\usepackage{microtype}      % microtypography
\usepackage{lipsum}
\usepackage{fancyhdr}       % header
\usepackage{graphicx}       % graphics
\graphicspath{{media/}}     % organize your images and other figures under media/ folder

\usepackage{amsmath}
\usepackage{algorithm}
\usepackage{algpseudocode}
\usepackage{dcolumn}% Align table columns on decimal point
\usepackage{bm}% bold math
\usepackage{mathptmx}
\usepackage{etoolbox}
\usepackage{float}
\usepackage{threeparttable}
\usepackage[table]{xcolor}
\usepackage[numbers]{natbib}
\usepackage{multirow}
\usepackage{lineno}
\usepackage{subcaption}
\usepackage{makecell}

\title{Data-Efficient Deep Operator Network for Unsteady Flow: A Multi-Fidelity Approach with Physics-Guided Subsampling}
\author{
  Sunwoong Yang \\
  Cho Chun Shik Graduate School of Mobility \\
  Korea Advanced Institute of Science and Technology (KAIST) \\
  Daejeon, 34051, Republic of Korea \\
  \texttt{sunwoongy@kaist.ac.kr} \\
  \texttt{\url{https://sites.google.com/view/aerodat}} \\
  \And
  Youngkyu Lee \\
  Division of Applied Mathematics \\
  Brown University \\
  182 George St, Providence, 02906 RI, USA \\
  \texttt{youngkyu\_lee@brown.edu} \\
  \texttt{\url{https://sites.google.com/view/youngkyulee}} \\
  \And
  Namwoo Kang \\
  Cho Chun Shik Graduate School of Mobility \\
  Korea Advanced Institute of Science and Technology (KAIST) \\
  Daejeon, 34051, Republic of Korea \\
  Also at Narnia Labs, Daejeon, 34051, Republic of Korea \\
  \texttt{nwkang@kaist.ac.kr} \\
  \texttt{\url{https://www.smartdesignlab.org/people/professor}} \\
}

\begin{document}
\maketitle

% \linenumbers
\begin{abstract}
This study presents an enhanced multi-fidelity deep operator network (DeepONet) framework for efficient spatio-temporal flow field prediction, with particular emphasis on practical scenarios where high-fidelity data is scarce. We introduce several key innovations to improve the framework's efficiency and accuracy. First, we enhance the DeepONet architecture by incorporating a merge network that enables more complex feature interactions between operator and spatio-temporal coordinates, achieving a 50.4\% reduction in prediction error compared to traditional dot-product operations. We further optimize the architecture through temporal positional encoding and point-based subsampling strategies, achieving a 7.57\% improvement in prediction accuracy while reducing training time by 96\%. Building upon this foundation, we develop a transfer learning-based multi-fidelity framework that leverages knowledge from pre-trained low-fidelity models to guide high-fidelity predictions. Our approach freezes the pre-trained branch and trunk networks while making only the merge network trainable during high-fidelity training, preserving valuable low-fidelity representations while efficiently adapting to high-fidelity features. Through systematic investigation, we demonstrate that this fine-tuning strategy not only outperforms linear probing and full-tuning alternatives but also surpasses conventional multi-fidelity frameworks by up to 76\%, while achieving up to 43.7\% improvement in prediction accuracy compared to single-fidelity training. As another core contribution, we introduce a physics-guided subsampling approach, which strategically selects high-fidelity training spatial points based on temporal dynamics identified by pre-trained low-fidelity models. This physics-guided subsampling strategy demonstrates remarkable effectiveness, achieving prediction accuracy comparable to a baseline model while reducing the high-fidelity sample requirement by 40\%. The robustness of these contributions, particularly our proposed MF-DeepONet framework and time-derivative subsampling, is further validated on a second, more complex dataset where they continue to demonstrate superior or highly competitive performance against conventional benchmarks. Through comprehensive experiments across multiple resolutions and two distinct datasets of increasing complexity, our enhanced framework demonstrates the potential to reduce the required high-fidelity dataset size while maintaining predictive accuracy.
\end{abstract}

% keywords can be removed

\section{Introduction}
\label{sec:intro}

In the era of scientific machine learning (SciML), accurate and efficient prediction of spatio-temporal flow fields plays a crucial role across various engineering disciplines, particularly in the development of digital twins that require real-time flow field predictions for virtual replicas of physical systems. While traditional numerical approaches like computational fluid dynamics (CFD) offer high accuracy, their computational demands make them impractical for real-time applications, where rapid prediction and flexible adaptation to different flow parameters are essential. This has led to the emergence of various deep learning approaches, including multilayer perceptrons (MLPs) \cite{wang2019non, yang2023inverse}, convolutional neural networks (CNNs) \cite{kang2022physics, maulik2021reduced, hasegawa2020machine, guastoni2021convolutional, hu2022mesh, eivazi2022towards}, and graph neural networks (GNNs) \cite{pfaff2020learning, yang2024towards, kim2024physics, yang2024enhancing}, as surrogate models for flow field prediction. However, these architectures face inherent limitations: MLPs lack spatial structure awareness and cannot flexibly handle varying mesh resolutions, requiring retraining for each new mesh configuration---a significant drawback for digital twins that must adapt to different system configurations. CNNs are constrained by fixed regular grid resolutions, limiting their applicability to irregular domains. GNNs, while capable of handling unstructured meshes, require extensive memory resources to store and process mesh connectivity information, making them computationally intensive for large-scale problems \cite{nabian2024x}.

The deep operator network (DeepONet) \cite{lu2021learning}, introduced to learn mappings between function spaces, offers a compelling alternative for spatio-temporal flow field prediction tasks by overcoming the limitations of traditional methods \cite{he2024predictions, bai2024data, lu2022comprehensive, taassob2024pinn, shukla2024comprehensive}. Its unique architecture, consisting of separate branch and trunk networks, enables prediction at arbitrary spatial locations without being constrained by specific grid resolutions or mesh connectivities. This resolution-invariant property makes DeepONet particularly suitable for flow field predictions where data may come from simulations with varying mesh configurations. However, the standard DeepONet architecture may not be ideally designed to capture the complex dynamics and intricate features of spatio-temporal behavior \cite{bai2024data, kontolati2024learning, karumuri2025physics, li2025architectural}. To fully exploit DeepONet's potential for spatio-temporal flow field prediction, we recognize the need for specialized architectural refinements and training methodologies tailored to fluid flow applications. By systematically developing and integrating enhancements to the core DeepONet framework, we aim to significantly improve both the accuracy and computational efficiency of flow field predictions through: 1) expanding the architecture with additional networks for more flexible feature interactions between operator and spatio-temporal coordinates, 2) addressing spectral bias through strategic positional encoding implementation, and 3) implementing efficient point-based subsampling techniques with automatic mixed precision training to reduce computational overhead.

However, training DeepONet remains challenging, especially when high-fidelity data is limited due to computational constraints. In this context, multi-fidelity modeling has emerged as a powerful approach in SciML, effectively leveraging relatively abundant low-fidelity data alongside limited high-fidelity data to enhance prediction accuracy while maintaining computational efficiency \cite{han2012hierarchical, meng2021fast, yang2024data, zhang2021multi, yang2022design, yang2022comment}. Existing multi-fidelity DeepONet frameworks \cite{lu2022multifidelity, de2023bi, demo2023deeponet} predominantly employ coupled architectures, where low-fidelity (LF) and high-fidelity (HF) networks are trained and utilized together. This coupling approach fundamentally requires identical query points for both branch inputs (function queries) and trunk inputs (spatio-temporal coordinates) across different fidelity levels. This severely restricts DeepONet's primary advantage: the flexibility to leverage operators at arbitrary points. It forces researchers to design low-fidelity and high-fidelity experiments with identical branch inputs and trunk inputs beforehand---a requirement that is highly inefficient and often impractical in industrial applications where data collection follows operational constraints rather than organized experimental design. While the multi-fidelity DeepONet proposed by \citet{howard2023multifidelity} successfully addressed this constraint by adopting a multi-fidelity deep neural network approach \cite{meng2020composite, zhang2021multi}, it still requires the coupled use of LF and HF networks during inference. This coupled architecture introduces computational overhead by requiring sequential two-network inference, which increases both memory requirements and computational costs. The inefficiency scales poorly with multiple fidelity levels, as each additional fidelity layer requires its own network, creating cascading overhead particularly problematic when integrated with physics-informed methods, where automatic differentiation should propagate through all architectures.

Due to this inherent coupling and the availability of matching LF and HF datasets, these approaches naturally adopt residual learning architectures. In these frameworks, LF DeepONet is trained specifically on LF data to produce baseline predictions, while a second separate network, residual DeepONet, is trained to model the discrepancy between the low-fidelity and high-fidelity outputs (i.e., $y_{HF}=y_{LF}+\hat{y}_{residual}$). More fundamentally, residual approaches assume that low-fidelity simulations capture essential physical phenomena with less accuracy---an assumption that can be challenging in fluid dynamics, where low-resolution simulations may struggle to fully represent critical structures like vortex formations or boundary layer effects. When LF and HF data differ in complex ways, residual DeepONet model must focus on modeling these complicated residual patterns, which is highly inefficient compared to directly predicting the HF values. This inefficiency stems from forcing the network to learn fundamentally missing physics rather than leveraging the strengths of end-to-end learning. The residual network must not only correct numerical inaccuracies but also compensate for entirely absent physical phenomena in the baseline LF DeepONet model, essentially requiring it to learn more complex mappings than a direct prediction approach would need.

Beyond these architectural limitations, fundamental methodological challenges exist in effectively leveraging the relationship between LF and HF data during DeepONet training. While \citet{xu2024multi} proposed a two-phase approach pre-training an LF DeepONet before training a residual DeepONet---which removes the requirement of identical LF and HF datasets by using its sequential transfer learning approach---their framework still necessitates two separate networks during inference stage. This decoupling between LF and HF DeepONets during training phase allows for more flexibility in dataset design, but continues to suffer from computational inefficiency as both models must be employed during inference. Additionally, it lacks a systematic mechanism to identify critical regions where high-fidelity training would be most beneficial, resulting in potentially inefficient use of the pre-trained low-fidelity model and valuable high-fidelity data. Combined with the inherent drawbacks of residual-based architectures discussed earlier, these issues highlight the need for a fundamentally different approach to multi-fidelity operator learning.

A more refined approach to multi-fidelity DeepONet for spatio-temporal flow field prediction would incorporate four key innovations: (1) a unified network structure that eliminates the need for coupled use of low-fidelity and high-fidelity DeepONets during inference, reducing computational overhead and memory requirements; (2) preservation of DeepONet's inherent query flexibility by removing the requirement for low-fidelity and high-fidelity datasets to share identical function inputs or spatio-temporal query points; (3) a physics-guided subsampling strategy that identifies regions of significant flow dynamics using pre-trained low-fidelity models, strategically focusing the training process on high-fidelity data from these dynamically critical areas to maximize information gain while minimizing computational cost; and (4) comprehensive evaluation of the effects of dataset fidelities and sizes on multi-fidelity model performance, with particular emphasis on strategies for minimizing high-fidelity data requirements while maintaining prediction accuracy.

To address these challenges comprehensively, we present a systematic investigation that directly tackles all four aspects through enhanced DeepONet architectures and efficient training strategies for spatio-temporal flow prediction, with a particular focus on developing practical multi-fidelity frameworks. Our key contributions can be organized into two main thrusts.

Our first step involves designing targeted enhancements to the DeepONet architecture for more accurate spatio-temporal flow field predictions:

\begin{enumerate}
    \item We propose an expanded DeepONet architecture incorporating a merge network for complex feature interactions between operator (branch net) and spatio-temporal coordinate spaces (trunk net). Through extensive experiments, we demonstrate that our element-wise multiplication merge strategy achieves a 50.4\% reduction in prediction error compared to traditional dot-product operations, while maintaining similar model parameters. This significant improvement highlights the importance of non-linear feature processing in capturing intricate flow dynamics.
    
    \item We systematically investigate positional encoding to establish practical guidelines for its application to spatio-temporal DeepONets. Our analysis reveals a crucial insight for physical systems: contrary to common practice in computer vision domains, encoding spatial coordinates degrades performance, while selective temporal encoding is beneficial, yielding a 7.57\% improvement in prediction accuracy. This provides a valuable, non-obvious guideline for optimizing coordinate representation in spatio-temporal operator learning.
    
    \item We leverage DeepONet's unique point-based prediction capability to develop efficient training protocols. Our approach demonstrates that training with a subset of spatial points significantly reduces computational overhead while maintaining prediction accuracy---challenging the assumption that more training points can lead to better results. Combined with automatic mixed precision training, we achieve a remarkable 96\% reduction in training time (from 8.672 to 0.350 hours) while maintaining or even enhancing prediction accuracy.
\end{enumerate}

Second, we develop and validate a novel multi-fidelity DeepONet (MF-DeepONet) framework that efficiently leverages both low- and high-fidelity data:

\begin{enumerate}
    \item We introduce a transfer learning-based MF-DeepONet framework that fundamentally departs from sequential, two-network approaches like those in \cite{lu2022multifidelity, de2023bi, demo2023deeponet}. By strategically transferring knowledge from pre-trained low-fidelity models to a single unified network for high-fidelity predictions, our approach eliminates both the computational overhead of maintaining coupled networks and the restrictive constraint of matching query points between fidelity levels: where the transfer learning strategy is inspired by previous works that have successfully applied transfer learning techniques to multi-fidelity modeling \cite{panigrahi2020survey, song2022transfer, chakraborty2021transfer, li2022line, de2022neural, jiang2023use, de2020transfer, aliakbari2022predicting}. This critical advancement preserves DeepONet's fundamental capability to evaluate operators at arbitrary points while removing the need for aligned branch and trunk inputs across fidelity levels---enabling seamless integration of heterogeneous data from different sources and resolutions. In direct comparison with conventional multi-fidelity approaches proposed by \citet{lu2022multifidelity}, our framework demonstrates superior performance with up to 76\% improvement in prediction accuracy while significantly reducing computational requirements during both training and inference, and the robustness of such performance is further validated on a second, more challenging high-resolution dataset.
    
    \item Through systematic investigation of different transfer learning configurations---including fine-tuning (where only the merge network is trainable while branch and trunk networks are frozen), full-tuning (where all network components are retrained), and linear probing (where only the final layer of the merge network is trainable)---we demonstrate that the selective fine-tuning approach significantly outperforms alternatives. Our extensive experimentation with various low-fidelity resolutions (16×16, 32×32, 64×64) and dataset sizes (50-300 samples) reveals that fine-tuning achieves up to 43.7\% improvement in prediction accuracy compared to single-fidelity training, while maintaining computational efficiency.
    
    \item We propose a physics-guided subsampling strategy based on temporal derivatives that efficiently identifies critical spatial areas requiring high-fidelity training. This intuitive approach demonstrates remarkable effectiveness: our method reduces prediction error by 10.02-20.73\% compared to traditional uniform sampling methods using the same amount of data. Furthermore, using only 60\% of high-fidelity samples with the same low-fidelity data, we achieve similar prediction accuracy compared to models trained without physics-guided sampling. The robustness of this physics-guided subsampling is also demonstrated on a more complex, higher-resolution dataset, where it remains an effective strategy that performs competitively against conventional residual-based sampling. These findings suggest a promising direction for reducing reliance on dense high-fidelity data in multi-fidelity modeling.
\end{enumerate}

The remainder of this paper is organized as follows. Section \ref{sec:method} presents the theoretical foundations of our enhanced DeepONet architecture and multi-fidelity framework, including the merge network design and time-derivative guided subsampling strategy. Section \ref{sec:data} describes the flow field data generation process with varying initial conditions and fidelity levels. Section \ref{sec:enhanced_deeponet} evaluates the architectural enhancements to DeepONet, analyzing the effects of different merge processes, positional encoding, and efficient training techniques. Section \ref{sec:MF} investigates our MF-DeepONet framework, comparing different transfer learning approaches and demonstrating performance with limited high-fidelity data. Section \ref{sec:TGS} explores the time-derivative guided subsampling strategy and quantifies its effectiveness in reducing high-fidelity data requirements. Section \ref{sec:newdata} further validates our framework on a second, more complex dataset and provides a comparative analysis of different guided sampling strategies. Finally, Section \ref{sec:conclusion} summarizes our findings and discusses implications for future research.

\section{Methodology}
\label{sec:method}

\subsection{DeepONet Architecture Enhancement}
\subsubsection{Enhanced DeepONet with Merge Network}\label{sec:method_merge}

The original DeepONet consists of two main components: a branch network and a trunk network. The branch network maps the input function information (e.g., initial conditions, boundary conditions, or other function characterizations) to a latent representation $\mathbf{B}(\mathbf{u}) \in \mathbb{R}^{p}$, where $\mathbf{u}$ is input function and $p$ is the dimension of the latent space. The trunk network takes spatio-temporal coordinates, $\mathbf{x}$; and it outputs latent space, $\mathbf{T}(\mathbf{x}) \in \mathbb{R}^{p}$. In the standard DeepONet formulation, these two latent representations are combined through a simple dot product operation to produce the final prediction: $f(\mathbf{u})(\mathbf{x})=\mathbf{B}(\mathbf{u}) \cdot \mathbf{T}(\mathbf{x})$ where function $f$ is an operator modeled by DeepONet.

For complex spatio-temporal flow field predictions, we hypothesize that this simple dot product operation may not provide sufficient non-linear modeling capacity. While \citet{lu2021learning} demonstrated that incorporating a bias term in the dot product slightly improves the model's performance in their applications, the effectiveness of this modification in complex flow field prediction remains to be verified. Therefore, we propose enhancing the architecture with a third component---a merge network---and investigate four strategies for merging branch and trunk network outputs, ranging from simple dot products to more complex non-linear operations through proposed merge networks:

\begin{itemize}
\item \textbf{Merge Type 0 (without bias)}: Simple dot product from the original DeepONet:
\begin{equation}
\text{Final output} = \mathbf{B}(\mathbf{u}) \cdot \mathbf{T}(\mathbf{x})
\end{equation}
\item \textbf{Merge Type 0 (with bias)}: Dot product with learnable bias for enhanced flexibility \cite{lu2021learning}:
\begin{equation}
\text{Final output} = \mathbf{B}(\mathbf{u}) \cdot \mathbf{T}(\mathbf{x}) + b
\end{equation}
\item \textbf{Merge Type 1}: Element-wise multiplication with additional non-linear processing by merge network:
\begin{equation}
\text{Final output} = \text{MergeNet}(\mathbf{B}(\mathbf{u}) \odot \mathbf{T}(\mathbf{x}))
\end{equation}
\item \textbf{Merge Type 2}: Concatenation with additional non-linear processing by merge network:
\begin{equation}
\text{Final output} = \text{MergeNet}([\mathbf{B}(\mathbf{u}); \mathbf{T}(\mathbf{x})])
\end{equation}
\end{itemize}

\begin{figure}[htb!]
    \centering
    \includegraphics[width=0.95\linewidth]{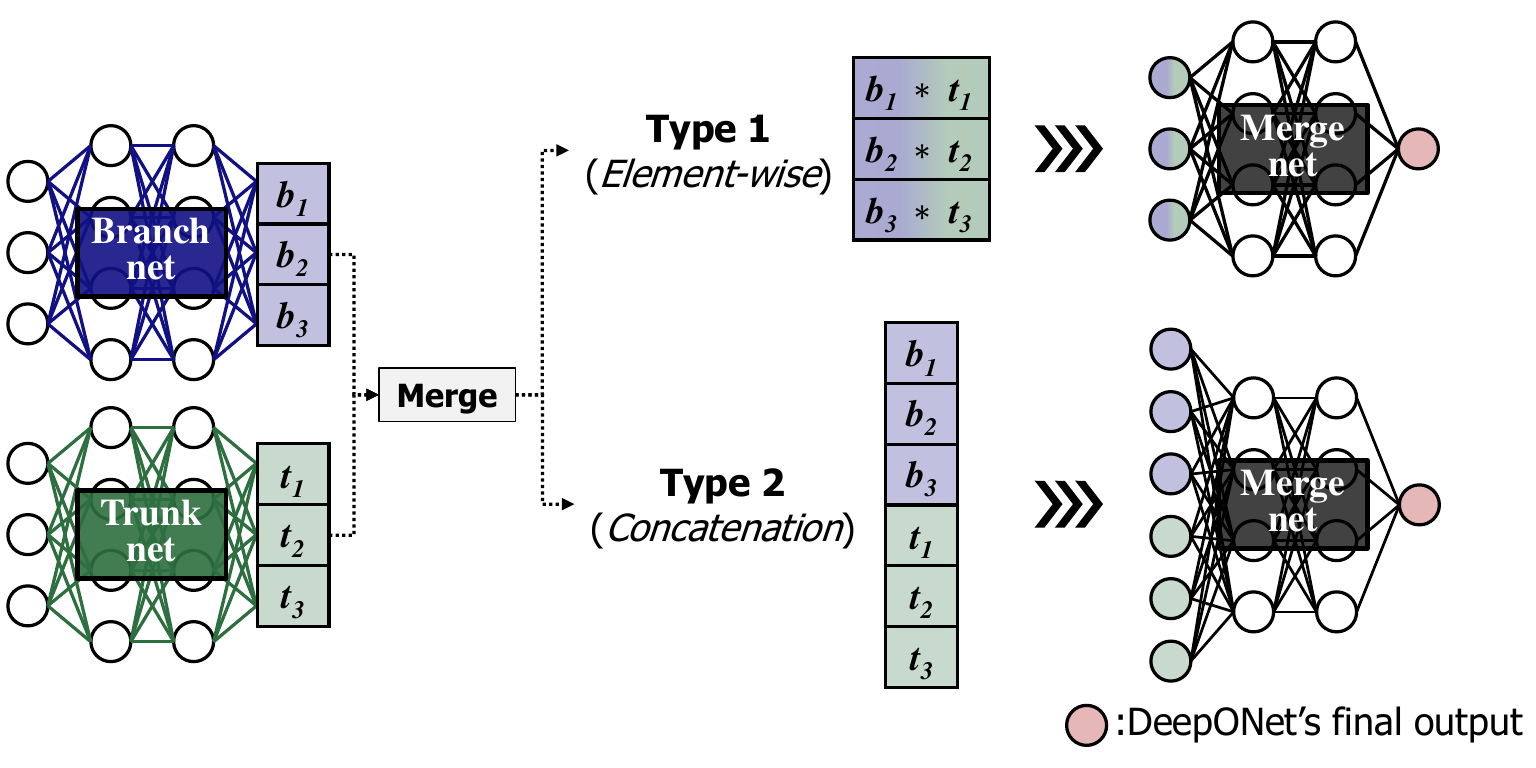}
    \caption{Schematic of two types of merging processes proposed in this study: Type 1 (element-wise) and Type 2 (concatenation).}
    \label{fig:MergeType}
\end{figure}

The introduction of MergeNet in Types 1 and 2, as visually represented in Fig.~\ref{fig:MergeType}, is intended to address a critical limitation of the original DeepONet architecture by enabling more complex non-linear interactions between branch and trunk network outputs. While the simple dot product operation in Type 0 may suffice for simpler operator learning tasks, more sophisticated non-linear processing of branch and trunk outputs can enhance complex spatio-temporal flow field predictions. The merge network provides this crucial non-linear modeling capacity by further processing the combined features through additional neural network layers, enabling the model to capture intricate relationships and dynamic patterns that may be missed by the basic dot product operation alone.

\subsubsection{Enhanced DeepONet with Positional Encoding}
\label{sec:method_PE}

Positional encoding (PE) has been widely used in various deep learning architectures, particularly in transformers and attention-based models, to incorporate spatial information into the input representations \cite{vaswani2017attention, mildenhall2021nerf}. In the context of DeepONet, the trunk net processes coordinate information which can suffer from spectral bias, limiting its ability to capture high-frequency components and intricate spatio-temporal dependencies \cite{wang2021understanding, tancik2020fourier}. By mapping coordinates to a higher-dimensional space beyond the original $(x,y,t)$ representation, PE is expected to enhance the network's ability to learn complex flow dynamics and temporal dependencies critical for accurate spatio-temporal flow field prediction.

\begin{figure}[htb!]
    \centering
    \includegraphics[width=0.8\linewidth]{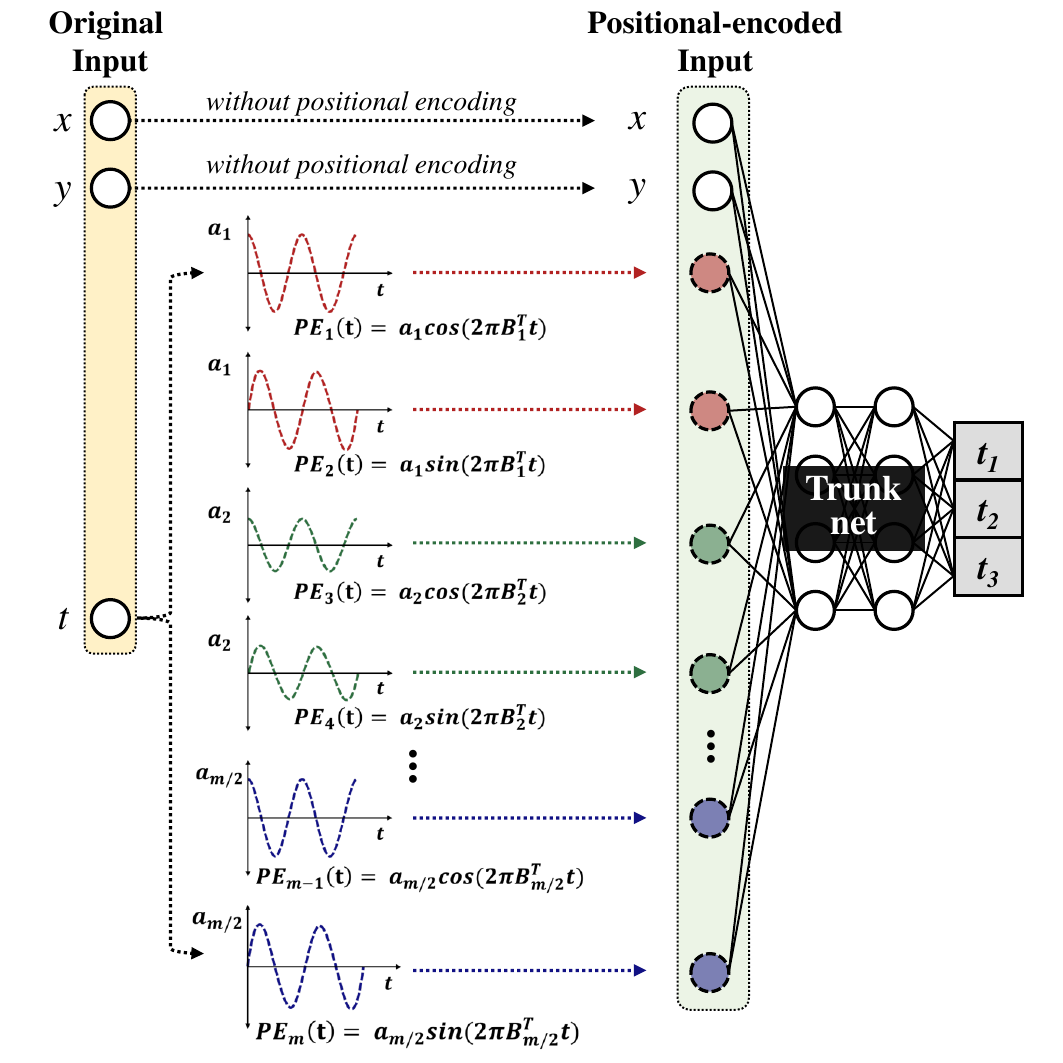}
    \caption{Schematic of the positional encoding process: this example shows the process where only temporal coordinate ($t$) is encoded.}
    \label{fig:PE}
\end{figure}

Specifically, when the PE transforms input coordinate $t$ to a vector $\gamma(t)$ using sinusoidal functions,
\begin{align}\label{eq:PE}
\gamma(t) = &[a_1\cos(2\pi B_1^Tt), a_1\sin(2\pi B_1^Tt), \ldots, a_{m/2}\cos(2\pi B_{m/2}^Tt), a_{m/2}\sin(2\pi B_{m/2}^Tt)]^T
\end{align}
where $B_j \in \mathbb{R}$ are frequency vectors sampled from a normal distribution and scaled by a factor $\sigma$, $a_j$ are amplitude factors (set to 1 in our implementation), and $m$ is the mapping size determining the dimension of the encoded features. This transformation, visualized in Figure~\ref{fig:PE}, maps the input to a higher-dimensional space, enabling better representation of high-frequency functions.

The PE implementation includes two key hyperparameters:
\begin{itemize}
\item \textbf{Scale Factor ($\sigma$)}: Controls the frequency range of the encoding, affecting the model's ability to capture different scales of variation.
\item \textbf{Mapping Size ($m$)}: Determines the dimension of the encoded features, controlling the expressiveness of the encoding.
\end{itemize}

Additionally, we introduce a flexible option that determines the scope of encoding:
\begin{itemize}
\item \textbf{PE($\mathbf{t}$)}: PE is applied only to the temporal coordinate in the trunk network input, preserving the original spatial coordinates.
\item \textbf{PE($\mathbf{x,y}$)}: PE is applied only to the spatial coordinates in the trunk network input, preserving the original temporal coordinate.
\item \textbf{PE($\mathbf{x,y,t}$)}: PE is applied to all coordinates (spatial and temporal) in the trunk network input, providing a uniform encoding scheme.
\end{itemize}

\subsection{Multi-Fidelity DeepONet Framework with Physics-Guided Subsampling Strategy}
\label{sec:method_MFDeepONet}

\subsubsection{Multi-Fidelity DeepONet Framework}
\label{sec:method_MFDeepONet2}

\begin{figure}[htb!]
    \centering
        \includegraphics[width=0.95\linewidth]{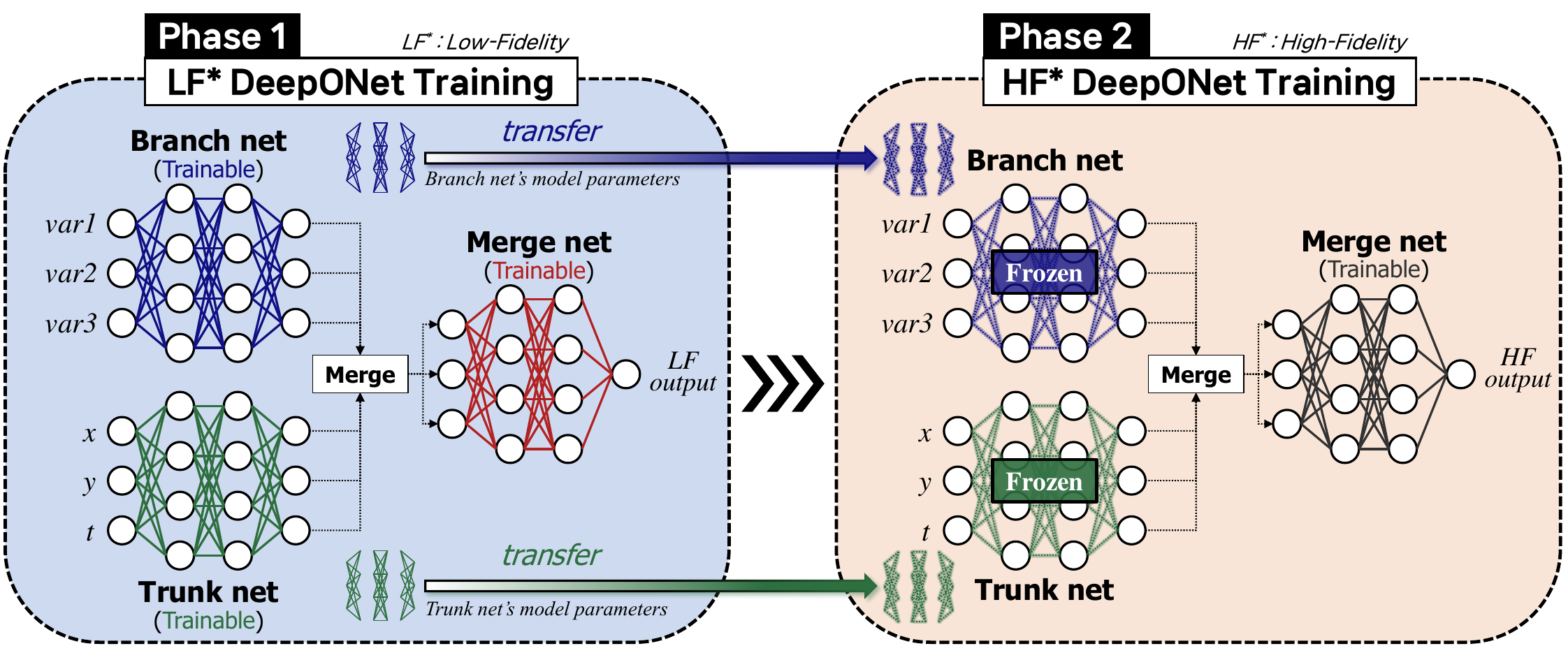}
        \caption{Two-phase training process of the MF-DeepONet framework. In \textbf{Phase 1}, all components are trained on low-fidelity data. In \textbf{Phase 2}, all networks are transferred but branch and trunk networks are frozen, while only the merge network is fine-tuned with high-fidelity data.}
        \label{fig:MFDeepONet_archit}
\end{figure}

We propose a multi-fidelity DeepONet (MF-DeepONet) framework that efficiently leverages both low-fidelity and high-fidelity data through knowledge transfer. Our two-phase approach first trains an LF DeepONet on relatively abundant low-fidelity data, then strategically transfers these pre-trained networks (LF DeepONet) to the high-fidelity model (HF DeepONet). By freezing the branch and trunk parameters of HF DeepONet (see Figure~\ref{fig:MFDeepONet_archit}), we preserve the fundamental flow physics captured during low-fidelity training. During the high-fidelity training phase, only the merge network remains trainable, allowing it to learn the specific refinements necessary for accurate high-fidelity predictions. Importantly, the merge network parameters transferred from Phase 1 serve as initial values that are further optimized using the high-fidelity dataset.

A key advantage of our approach is its computational efficiency during inference. While training involves two distinct phases, once training is complete, only the final HF DeepONet model in Phase 2 is needed for predictions. This creates a significant computational advantage compared to conventional coupled multi-fidelity frameworks that require both low and high-fidelity networks to be executed in sequence during inference. Our decoupled design effectively eliminates this computational overhead, resulting in substantially faster inference times and reduced memory requirements during deployment---critical factors for real-time applications and resource-constrained environments. Furthermore, this decoupled architecture enables the use of flexible LF and HF datasets, as it doesn't require corresponding data pairs between fidelity levels.

To comprehensively understand the impact of different knowledge transfer strategies, we will explore three distinct transfer learning approaches within our framework. While our primary proposed MF-DeepONet strategy employs fine-tuning (where only the merge network is trainable while branch and trunk networks are frozen), we additionally investigate \cite{kumar2022fine}: (1) full-tuning, where all network components (branch, trunk, and merge) of HF DeepONet are trainable, allowing complete adaptation to high-fidelity data; and (2) linear probing, the most restrictive approach that freezes all layers except the final layer of the merge network.

\subsubsection{Physics-Guided Subsampling Strategy}
\label{sec:time_sampling}

After training the LF DeepONet model, we aim to efficiently utilize the available HF dataset for training the HF DeepONet. While DeepONet's point-based architecture already offers flexibility in training with either complete or subsampled spatial points (note that effects of subsampling will be investigated in Section 4.2), we propose further enhancing this capability through physics-guided subsampling. Specifically, we introduce a time-derivative subsampling technique that strategically identifies dynamically important regions in the flow field. This approach leverages the temporal derivative information from the pre-trained LF model to guide the selection of spatial query points from the HF dataset, potentially enabling more efficient training of HF DeepONet by focusing computational resources on regions with significant flow dynamics.

\begin{figure}[htb!]
\centering
\includegraphics[width=0.6\textwidth]{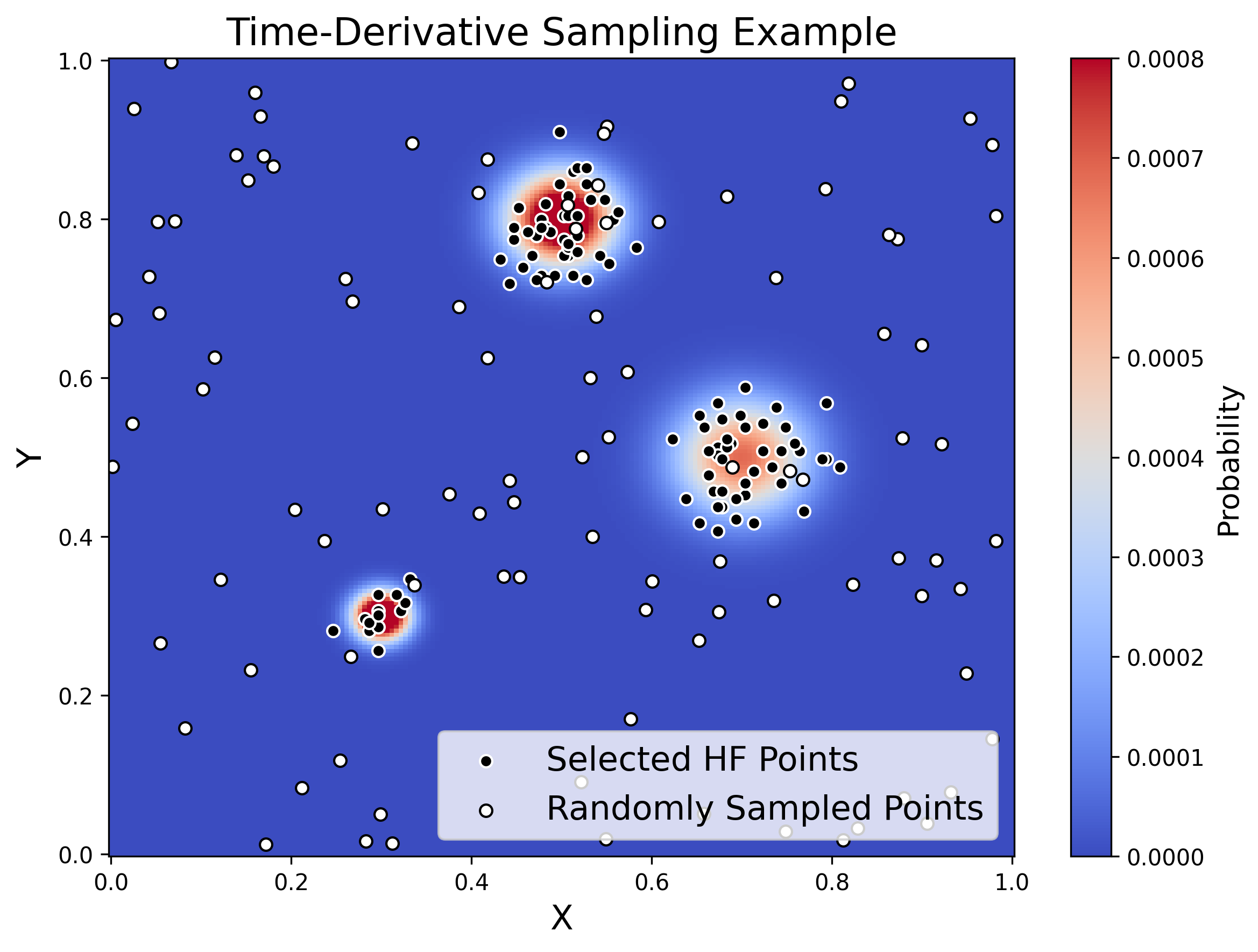}
\caption{Time-derivative guided subsampling demonstration: background shows probability distribution based on temporal derivatives, with black dots representing points selected using this distribution and white circles showing randomly sampled points for spatial coverage.}
\label{fig:sampling}
\end{figure}

The complete procedure for selecting HF training points using time-derivative guided subsampling is formally described in Algorithm~\ref{alg:time_deriv_sampling}. This physics-guided approach first computes the temporal derivatives at each spatial point from the HF dataset using the pre-trained LF model through automatic differentiation (lines 4-6). For each HF spatial location, we calculate a score based on the average magnitude of temporal derivatives predicted by the LF model (lines 7-9). Points with larger temporal derivatives---indicating regions of significant flow dynamics---receive higher sampling probabilities after score transformation and normalization (lines 10-11). Rather than deterministically selecting the highest-scoring points, we employ probabilistic sampling using probability defined by these score values, which introduces beneficial stochasticity while still favoring dynamically important regions. The final training set of $N_{subsampled}$ points is constructed by first selecting a fraction ($N_{subsampled} \cdot r$) via the probabilistic approach (line 13), and then sampling the remaining points ($N_{subsampled} \cdot (1-r)$) uniformly to ensure spatial coverage (line 14). By strategically focusing computational resources on dynamically important regions while preserving randomness, this approach can achieve superior prediction accuracy by using a more informative, smaller subset of HF spatial training points than a purely random sampling would. Figure~\ref{fig:sampling} visually demonstrates our time-derivative guided subsampling strategy, showing how high-fidelity training points are strategically selected based on a combination of derivative-weighted probability distributions and uniform random sampling to maintain spatial coverage.

\begin{algorithm}[htb!]
\caption{Time-derivative guided subsampling strategy}
\label{alg:time_deriv_sampling}
\begin{algorithmic}[1]
\State \textbf{Input:} HF dataset spatial coordinates $\mathbf{x}_{HF}$, pre-trained LF model, sampling ratio $r$
\State \textbf{Output:} Selected spatial points for HF DeepONet training
\State \textbf{Note:} $\hat{\omega}^{LF}(\mathbf{x}_{HF})$ denotes the vorticity field predicted by the LF model at coordinates $\mathbf{x}_{HF}$
\For{each spatial point $\mathbf{x}_{HF}$}
    \State Compute $\frac{\partial \hat{\omega}^{LF}(\mathbf{x}_{HF})}{\partial t}$ using automatic differentiation
\EndFor
\For{each spatial point $\mathbf{x}_{HF}$}
    \State $\text{score}(\mathbf{x}_{HF}) = \frac{1}{T} \sum_{t=1}^T |\frac{\partial \hat{\omega}^{LF}(\mathbf{x}_{HF})}{\partial t}|$
\EndFor
\State Transform scores using power-law: $p(\mathbf{x}_{HF}) = \text{score}^2(\mathbf{x}_{HF})$
\State Normalize: $p(\mathbf{x}_{HF}) = \frac{p(\mathbf{x}_{HF})}{\sum p(\mathbf{x}_{HF})}$
\State To subsample $N_{subsampled}$ spatial points from the HF dataset for training HF DeepONet:
\State \quad Sample $N_{subsampled} \cdot r$ points using multinomial distribution with probabilities $p(\mathbf{x}_{HF})$
\State \quad Sample remaining $N_{subsampled} \cdot (1-r)$ points uniformly at random
\end{algorithmic}
\end{algorithm}

\subsubsection{Key Considerations for Effective Multi-Fidelity Modeling}
\label{sec:limit-MF}

Our MF-DeepONet framework with physics-guided subsampling fundamentally relies on the predictive capability of the pre-trained LF DeepONet, meaning its effectiveness is inherently tied to the quality and computational efficiency of the low-fidelity dataset. Therefore, a critical preliminary step in applying any MF strategy, including ours, is the careful selection and validation of the LF model. This selection process must consider two key aspects: (1) the degree to which the LF model can capture the essential physical phenomena relevant to the HF model, as this dictates the potential for effective knowledge transfer, and (2) the computational cost of generating LF data, since an MF approach achieves its overall efficiency gain primarily when LF simulations are meaningfully less computationally expensive than HF simulations. If the LF data are not significantly less expensive to generate, or if they fail to adequately represent the critical physics of the HF data, the advantages of the multi-fidelity approach are naturally diminished. However, this limitation is not unique to our framework—it represents a common challenge shared across multi-fidelity modeling strategies that incorporate LF data, whether explicitly or implicitly.

\section{Data Generation: Flow Fields with Varying Initial Conditions}
\label{sec:data}

The training datasets comprise both LF and HF flow simulations of incompressible Navier-Stokes equations in a two-dimensional periodic domain $[0
,1m] \times [0,1m]$. The numerical solution is obtained using a pseudo-spectral solver that employs fast Fourier transforms for spatial discretization, with a 2/3 dealiasing rule to prevent aliasing errors. The solver implements a semi-implicit time integration scheme where advection terms are treated explicitly while diffusion terms are handled implicitly.

\begin{figure}[htb!]
    \centering
    % First (left) subplot
    \begin{subfigure}[b]{0.24\linewidth}
        \centering
        \includegraphics[width=\linewidth]{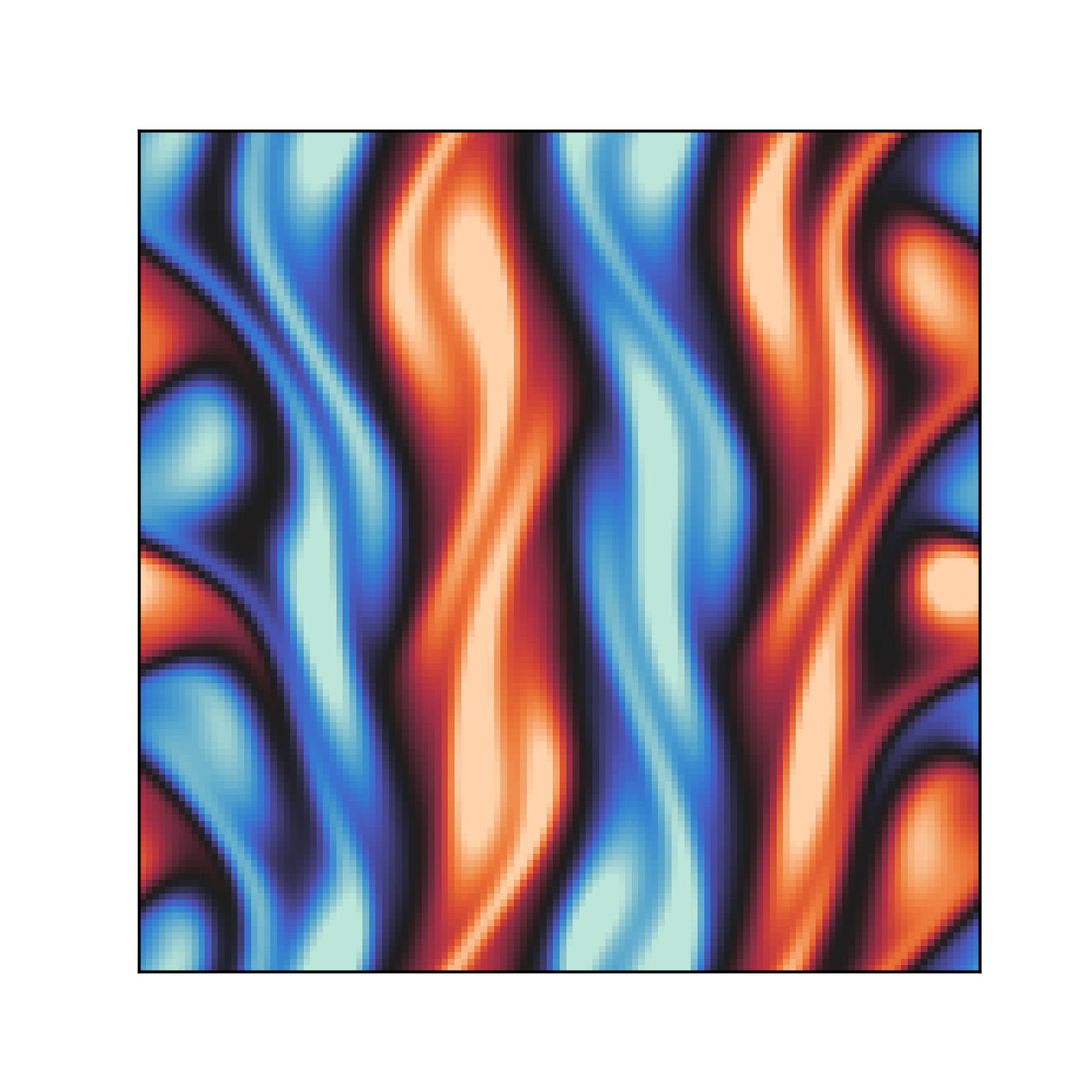}
        \caption{Example 1}
        \label{fig:diverse_ICs1}
    \end{subfigure}%
    % Second (right) subplot
    \begin{subfigure}[b]{0.24\linewidth}
        \centering
        \includegraphics[width=\linewidth]{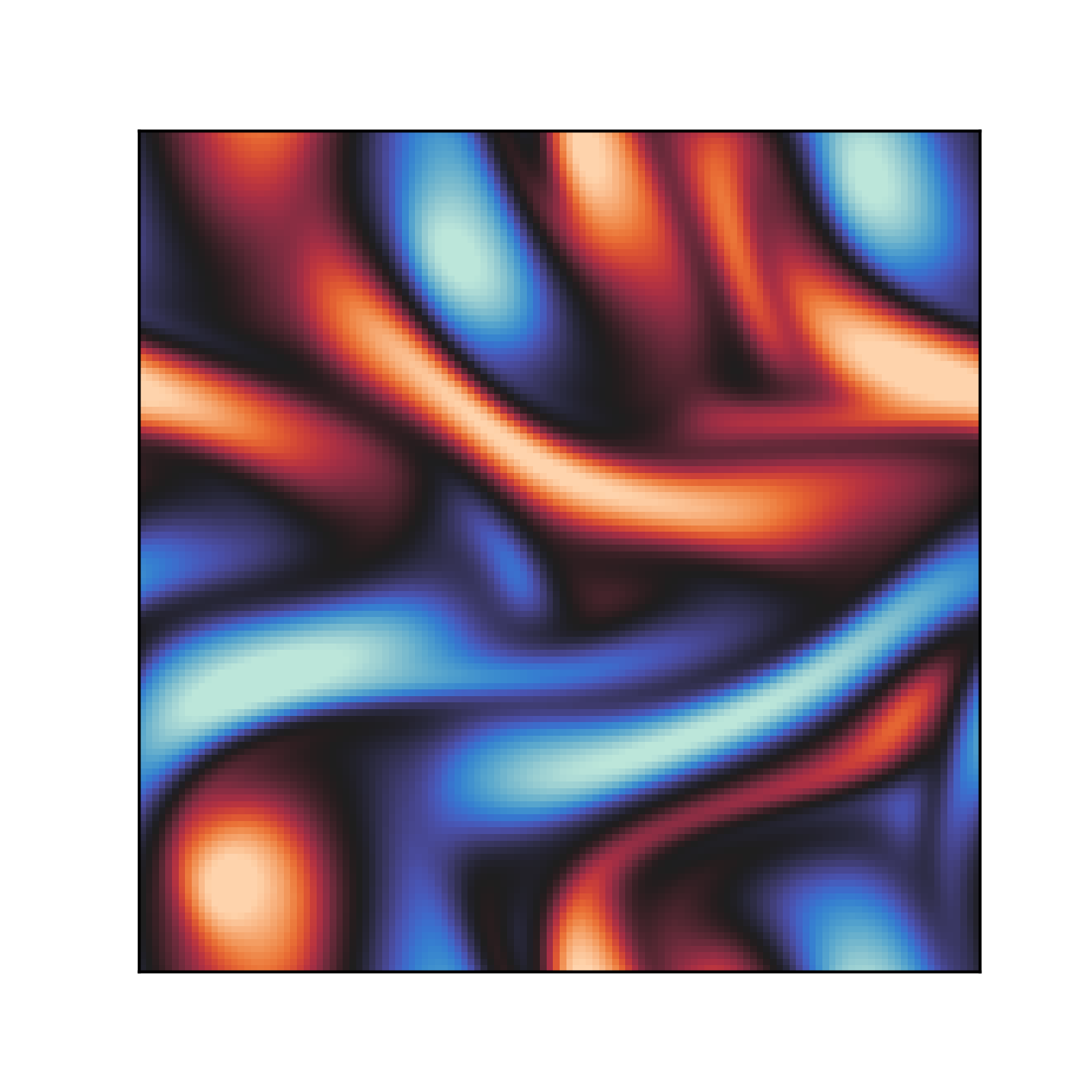}
        \caption{Example 2}
        \label{fig:diverse_ICs2}
    \end{subfigure}%
    \begin{subfigure}[b]{0.24\linewidth}
        \centering
        \includegraphics[width=\linewidth]{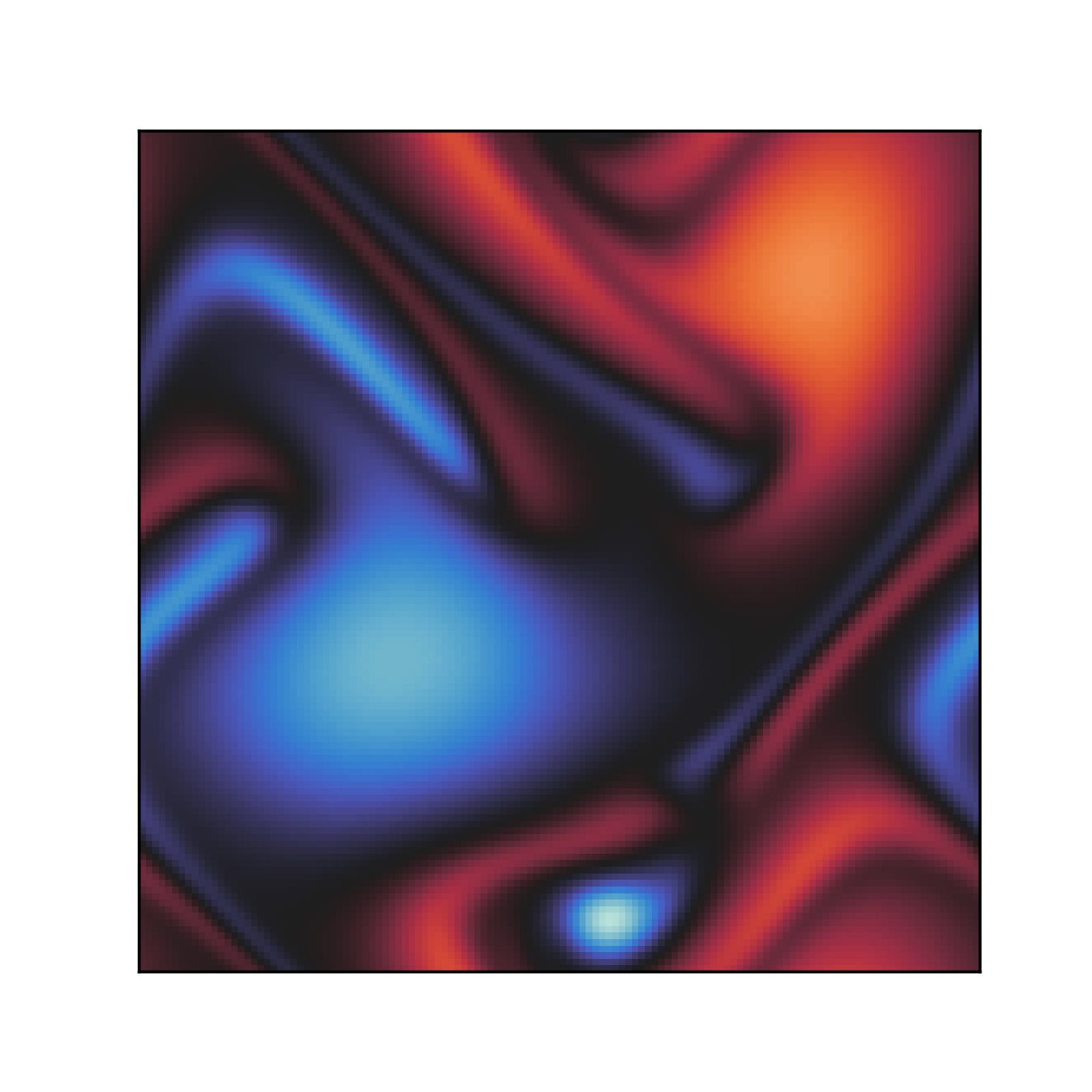}
        \caption{Example 3}
        \label{fig:diverse_ICs3}
    \end{subfigure}%
    \begin{subfigure}[b]{0.24\linewidth}
        \centering
        \includegraphics[width=\linewidth]{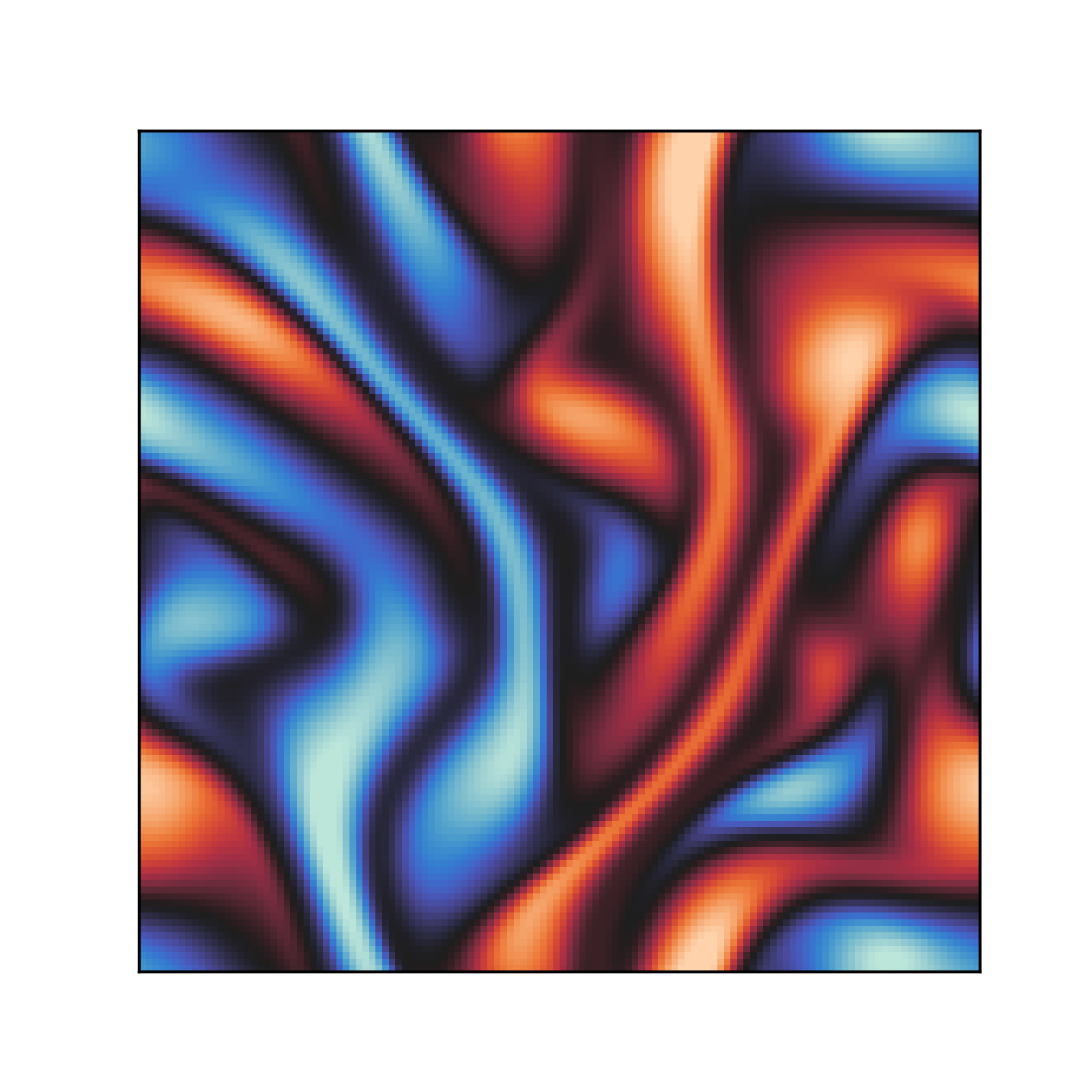}
        \caption{Example 4}
        \label{fig:diverse_ICs4}
    \end{subfigure}
    \caption{Vorticity fields at $t=0.5s$ for four distinct initial conditions, each defined by a different combination of $var_1$ and $var_2$. This diversity in the observed patterns highlights the complexity and challenges of predicting the flow evolution.}
    \label{fig:diverse_ICs}
\end{figure}

First, for the HF dataset, we generate 100 training samples and 50 test samples on a uniform grid of 128×128 resolution. Each sample is initialized with different combinations of two parameters ($var_1$, $var_2$) that define the initial velocity field components:
\begin{equation}\label{eq:ICs}
\begin{aligned}
v_x(x,y,t=0) &= -\sin(var_1 \pi y) \\
v_y(x,y,t=0) &= \sin(var_2 \pi x)
\end{aligned}
\end{equation}
where $var_1$ and $var_2$ are randomly sampled from uniform distributions in the range $\mathcal{U}[1,3]$. The Reynolds number is fixed at $Re=1,000$ for all simulations. The temporal evolution of the flow field is computed with a time step of $\Delta t = 2.5 \times 10^{-5} [s]$, solving until $t=1s$. While the solver operates at this fine temporal resolution, the flow field data is extracted at a coarser interval of $\Delta t_{train} = 5 \times 10^{-3} [s]$ for training purposes. This output frequency is selected as it adequately captures relevant flow dynamics while significantly reducing the number of temporal snapshots. This approach enhances computational efficiency for the subsequent AI model training, which can effectively learn the continuous dynamics from this sparser data, thus reducing overall training overhead. Importantly, these simulations feature no external forcing after $t=0$, resulting in gradually dissipating flow fields where vorticity magnitude decreases over time. This dissipative behavior creates challenging temporal dynamics that are particularly difficult to predict accurately as the flow evolves further from its initial state. Throughout this study, we focus on predicting the vorticity field, a derived quantity that captures the local rotation in the fluid and presents a more challenging prediction target than primary variables like velocity components. Figure~\ref{fig:diverse_ICs} displays the vorticity field at $t=0.5s$ for four distinct initial conditions. The marked diversity in the flow patterns, resulting from variations in $var_1$ and $var_2$, underscores the complex nature of the dataset.

\begin{figure}[htb!]
\centering
\includegraphics[width=0.65\textwidth]{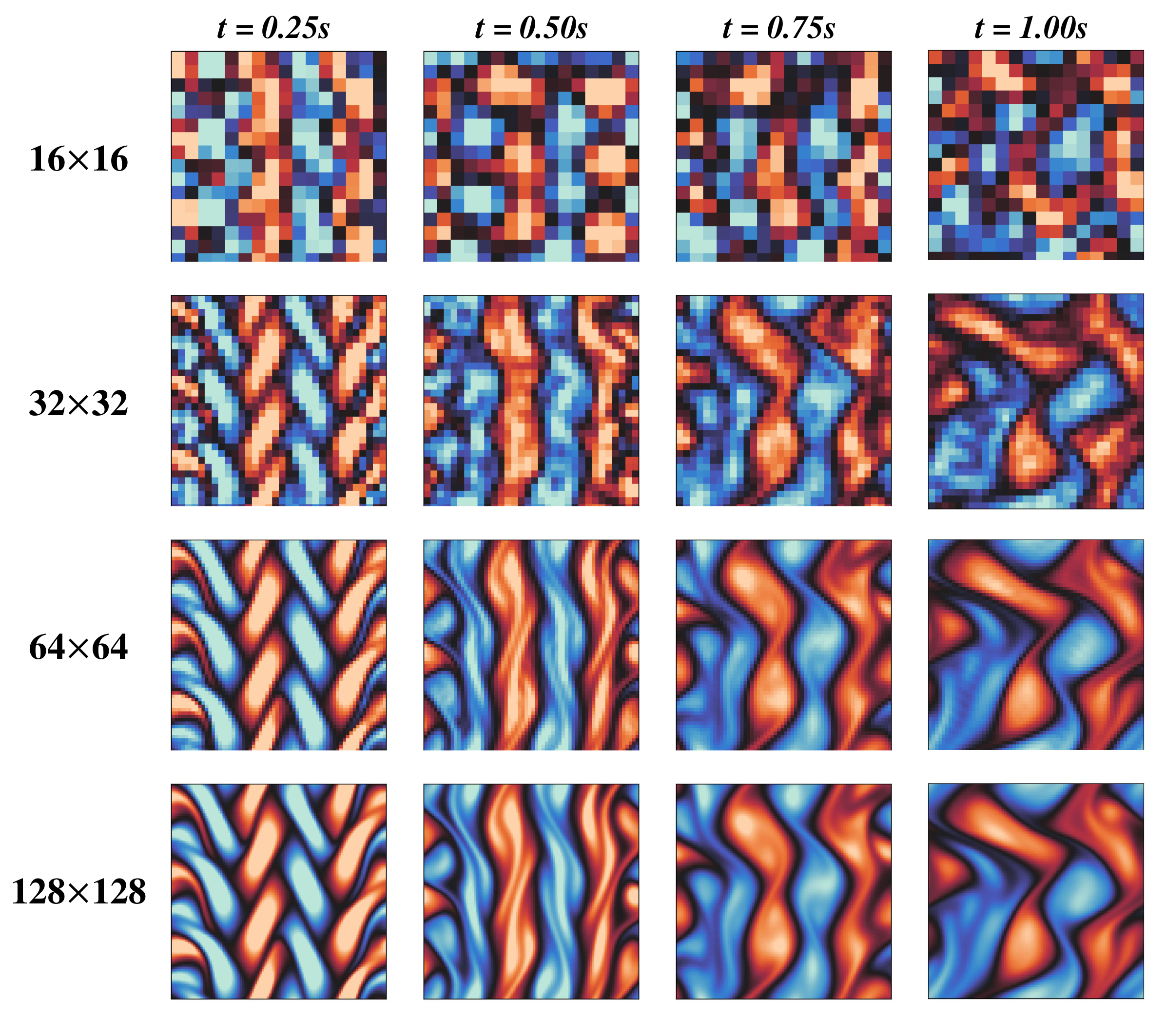}
\caption{Effects of grid resolution on vorticity field evolution with fixed parameters $var_1=3$ and $var_2=1$ (contours represent vorticity values). It also shows how dissipative effects intensify over time in the adopted flow field dataset.}
\label{fig:data1}
\end{figure}

For the LF datasets, we generate 300 training samples for each of three different grid resolutions: 16×16, 32×32, and 64×64. While the generation of HF 128×128 data requires approximately 360 seconds per sample using AMD EPYC 7282 processor, the LF data require 90, 100, and 140 seconds for 16×16, 32×32, and 64×64 resolutions, respectively. In total, generating the HF dataset (100 training samples) requires 10 hours, whereas generating each LF dataset (300 samples) requires 7.5, 8.3, and 11.7 hours, respectively. As illustrated in Figure~\ref{fig:data1}, the grid resolution significantly impacts the accuracy of flow field predictions. The coarsest resolution (16×16) shows severe limitations in capturing flow evolution: while it approximates the macroscopic flow patterns at $t=0.25s$, the flow field becomes notably distorted from $t=0.5s$ onward compared to higher-resolution simulations. This suggests that data from such coarse resolutions might be inadequate or even detrimental in a multi-fidelity framework. However, from 32×32 resolution upward, the simulations begin to capture the essential macroscopic flow properties visible in the HF (128×128) case. This observation suggests that LF data from these intermediate resolutions (32×32 and 64×64) could effectively support HF predictions in our multi-fidelity framework.

\textbf{\textit{Note on dataset nesting in multi-fidelity modeling}}: Although the LF datasets used in this study are spatially nested within the HF datasets---i.e., their grid points are subsets of the higher-resolution grids (e.g., 64×64 is fully contained within 128×128)---the input function parameters ($var_1$ and $var_2$) are sampled independently across fidelities. Consequently, the datasets are not nested in terms of input function parameters. This is a critical distinction because multi-fidelity strategies often assume corresponding LF and HF outputs exist for identical input queries \cite{lu2022multifidelity, de2023bi, demo2023deeponet}, which is not the case here due to independent parameter sampling across fidelities. This design choice aligns with our objective to develop an MF-DeepONet framework that operates robustly without requiring paired or nested datasets, thereby enhancing its practical applicability in real-world scenarios where consistent parameter sampling across fidelities is often infeasible. It should also be noted that our framework can accommodate datasets with non-nested spatio-temporal grid points, which is not explored in this study.

\section{Improvement of DeepONet Architecture for Spatio-Temporal Flow Field Prediction}\label{sec:enhanced_deeponet}

In this section, we focus on enhancing single-fidelity DeepONet performance using only high-fidelity data (128×128 resolution). We systematically investigate the effects of different merge processes, spatial point subsampling approaches, and positional encoding strategies to establish the most effective base architecture. This optimized DeepONet architecture will serve as the foundation for both LF and HF models in our multi-fidelity framework presented later.

\subsection{Effects of Different Merge Process of Branch/Trunk Nets}\label{sec:pre_merge} 

We first investigate the effectiveness of different strategies for combining outputs from the branch and trunk networks elaborated in Section~\ref{sec:method_merge}. To ensure a fair comparison, we carefully adjust the network architectures to maintain similar numbers of learnable parameters across different configurations (approximately 300K parameters). For the basic dot-product variants, Merge Type 0 with and without bias, we employ more complex branch and trunk networks to match the parameter count of models with Type 1 and 2 networks. Additionally, we implement the modified DeepONet architecture proposed by \citet{wang2022improved}, which introduces encoder layers for enhanced nonlinear interactions but still relies on the fundamental dot-product mechanism between branch and trunk networks at its core. Across all DeepONet configurations, the branch network consistently takes two input parameters ($var_1$ and $var_2$) defining initial conditions, while the trunk network processes three spatial-temporal coordinates ($x$, $y$, and $t$). For detailed network architectures, please refer to Table~\ref{tab:merge_comparison}. All models are trained for 800 epochs using the Adam optimizer with an initial learning rate of $10^{-3}$, GeLU activation functions, and minibatch size of 1 on an NVIDIA A6000 GPU. To account for training stochasticity and ensure reproducible results, all experiments in this paper are repeated at least three times with different random seeds, and we report the averaged values of these runs. To evaluate the predictive accuracy of each model, we report the test error as the root mean squared error (RMSE) computed across all spatio-temporal test points and test cases. Specifically, for a set of $N_{\text{test}}$ test cases, each with $N_t$ temporal snapshots and $N_x$ spatial points, the RMSE is defined as:

\begin{equation}
\text{RMSE} = \sqrt{
\frac{1}{N_{\text{test}} \cdot N_t \cdot N_x}
\sum_{i=1}^{N_{\text{test}}}
\sum_{j=1}^{N_t}
\sum_{k=1}^{N_x}
\left( \hat{\omega}_{ijk} - \omega_{ijk} \right)^2
}
\end{equation}

where $\hat{\omega}_{ijk}$ and $\omega_{ijk}$ denote the predicted and ground truth vorticity values, respectively, for the $i$-th test case, at time $t_j$, and spatial location $\mathbf{x}_k$. All test errors reported in this paper, including those in Table~\ref{tab:merge_comparison}, are expressed as these RMSE values: this definition is expected to ensure that the reported test error reflects a comprehensive average across all spatio-temporal predictions in the test set, providing a physically meaningful and consistent measure of overall model accuracy.

\begin{table}[htb!]
\centering
\setlength{\abovecaptionskip}{10pt} % Spacing
\renewcommand{\arraystretch}{1.1}
\caption{Comparison of different merge strategies for DeepONet}
\label{tab:merge_comparison}
% \begin{tabular}{l@{\hspace{0.5cm}}cccc}
\begin{tabular}{lcccc}
    \hline
    Merge Strategy & Network Architecture & Parameters & Training Hour & Test Error \\ \hline
    Type 0 (Dot-Product w/o bias) & \begin{tabular}{@{}c@{}} Branch: 2-128×6 \\ Trunk: 3-64-128×6 \end{tabular} & 306,304 & 6.197 & 34.447 \\ \hline
    Type 0 (Dot-Product w/ bias) & \begin{tabular}{@{}c@{}} Branch: 2-128×6 \\ Trunk: 3-64-128×6 \end{tabular} & 306,305 & 6.453 & 34.546 \\ \hline
    \textbf{Type 1 (Element-wise)} & \begin{tabular}{@{}c@{}} Branch: 2-128 \\ Trunk: 3-64-128 \\ Merge: 128×6-64-32-16-1 \end{tabular} & 300,289 & 8.672 & \textbf{17.079} \\ \hline
    Type 2 (Concatenation) & \begin{tabular}{@{}c@{}} Branch: 2-128 \\ Trunk: 3-64-128 \\ Merge: 256-128×5-64-32-16-1 \end{tabular} & 292,033 & 8.399 & 23.060 \\ \hline
    \citet{wang2022improved} & \begin{tabular}{@{}c@{}} Branch: 2-128 \\ Trunk: 3-64-128 \\ Merge: 256-128×5-64-32-16-1 \end{tabular} & 302,336 & 17.918 & 32.099 \\ \hline
\end{tabular}
\end{table}

The results in Table~\ref{tab:merge_comparison} demonstrate three key findings. First, incorporating merge networks significantly improves prediction accuracy compared to the two Type 0 approaches suggested by \citet{lu2021learning}. While the original dot-product architecture was successful for simpler operator learning tasks in other studies, our results show its limitations in complex flow field predictions: the test error with dot-product operations (both with and without bias) remains above 34, indicating substantial prediction inaccuracies. In contrast, the introduction of merge networks dramatically reduces the test error to 17-23, despite requiring longer training times (approximately 8.5 hours compared to 6.2 hours) due to the additional processing of branch and trunk outputs through the distinct merge network. This stark improvement in accuracy---reducing test error from 34.447 to 17.079, a 50.4\% improvement---highlights that additional non-linear processing of combined features is crucial for capturing complex spatio-temporal flow dynamics. The merge network effectively acts as a learnable non-linear mapping that can adapt to intricate relationships between the latent representations from branch and trunk networks, a capability that simple dot-product operations alone cannot provide.

Second, the comparison with \citet{wang2022improved}'s approach provides additional evidence for the limitations of dot-product mechanisms in complex flow field predictions. Despite having a similar parameter count (302,336 parameters), \citet{wang2022improved}'s method achieves a test error of 32.099, which is substantially higher than our Type 1 (17.079) and Type 2 (23.060) approaches. Moreover, this approach requires significantly longer training time (17.918 hours) compared to our Type 1 (8.672 hours) and Type 2 (8.399 hours) methods, indicating computational inefficiency due to its additional architectural complexity. Notably, while this approach outperforms the basic dot-product operations (Type 0), it still falls short of our proposed merge network strategies in both accuracy and computational efficiency.

Third, among the enhanced architectures, Type 1 (element-wise multiplication) outperforms Type 2 (concatenation), achieving a test error of 17.079 compared to 23.060. This advantage stems from the structured interaction in Type 1, which more naturally aligns with well-established techniques such as modal decomposition (e.g., proper orthogonal decomposition) and spectral expansions, both of which have demonstrated strong performance in CFD.

\begin{enumerate}

    \item \textbf{Proper Orthogonal Decomposition (POD) Perspective:}
    Element-wise multiplication in Type 1 architectures directly mirrors the fundamental structure of POD, where complex flow fields are reconstructed through coefficients multiplied by basis modes \cite{taira2017modal}. In DeepONet, this correspondence is explicit: the branch network generates coefficients, while the trunk network produces basis functions---creating a natural factorization that resembles POD's coefficient-times-basis-function formulation. This structural alignment with POD, which has proven highly effective in CFD for reduced-order modeling and flow reconstruction \cite{eivazi2022towards, kang2022physics}, explains why Type 1's element-wise multiplication outperforms Type 2's less structured concatenation approach.

    \item \textbf{Spectral Methods Perspective:} 
    Type 1 approach also aligns with spectral methods, where solutions are represented as weighted sums of basis functions \cite{li2025architectural}. Here, the trunk network acts as a global basis generator, while the branch network provides coefficients---akin to spectral expansions in Fourier or Laplace methods. Since spectral methods have already demonstrated strong performance in solving the Navier-Stokes equations \cite{canuto1998stabilized, spalart1991spectral}, enforcing this structured decomposition allows Type 1 to better capture the underlying physics of flow evolution, outperforming the less structured feature fusion in Type 2.

\end{enumerate}

\begin{figure}[htb!]
    \centering
    % First (left) subplot
    \begin{subfigure}[b]{0.49\linewidth}
        \centering
        \includegraphics[width=\linewidth]{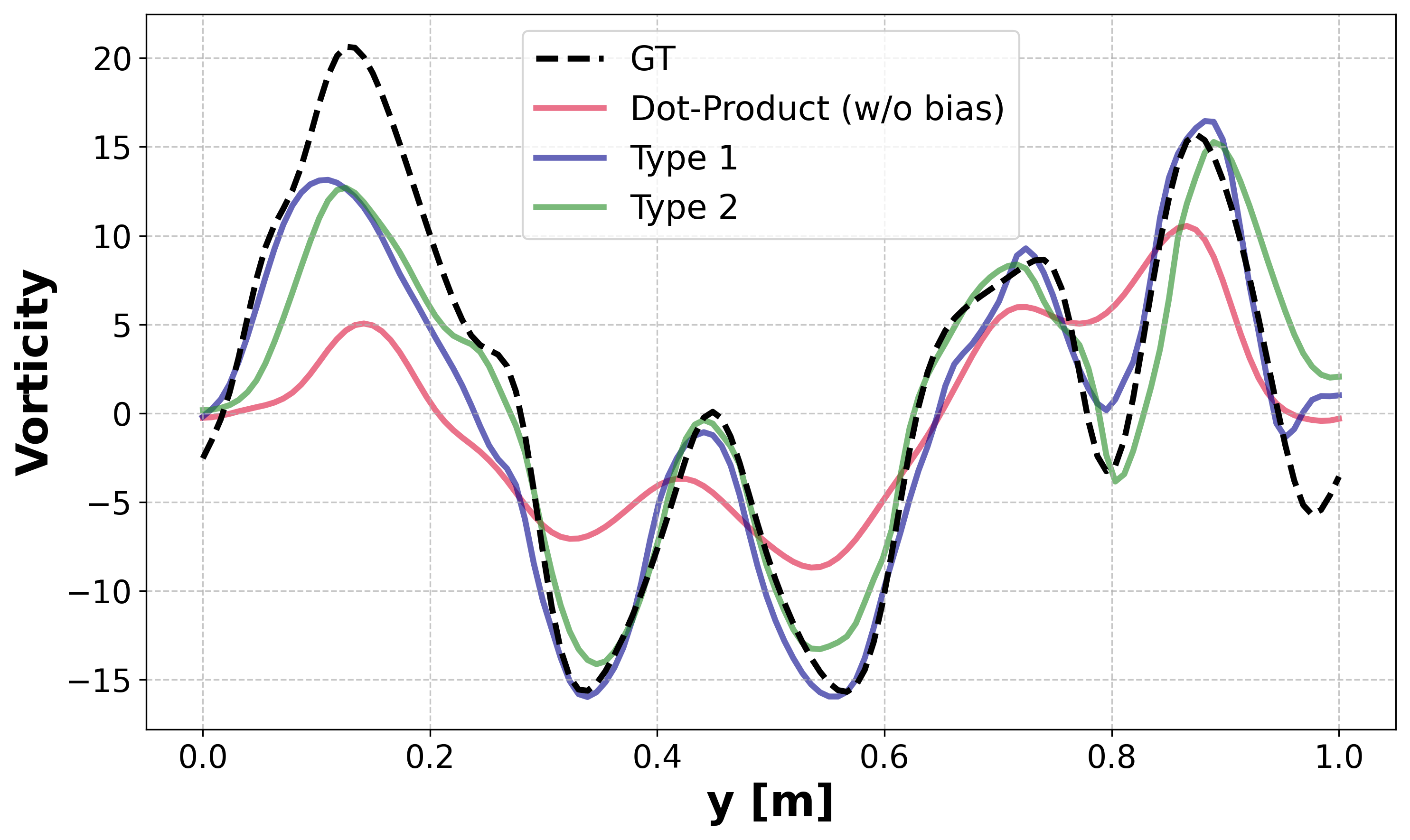}
        \caption{$t=0.5s$}
        \label{fig:merge_Sampling_lineplot1}
    \end{subfigure}
    % Second (right) subplot
    \begin{subfigure}[b]{0.49\linewidth}
        \centering
        \includegraphics[width=\linewidth]{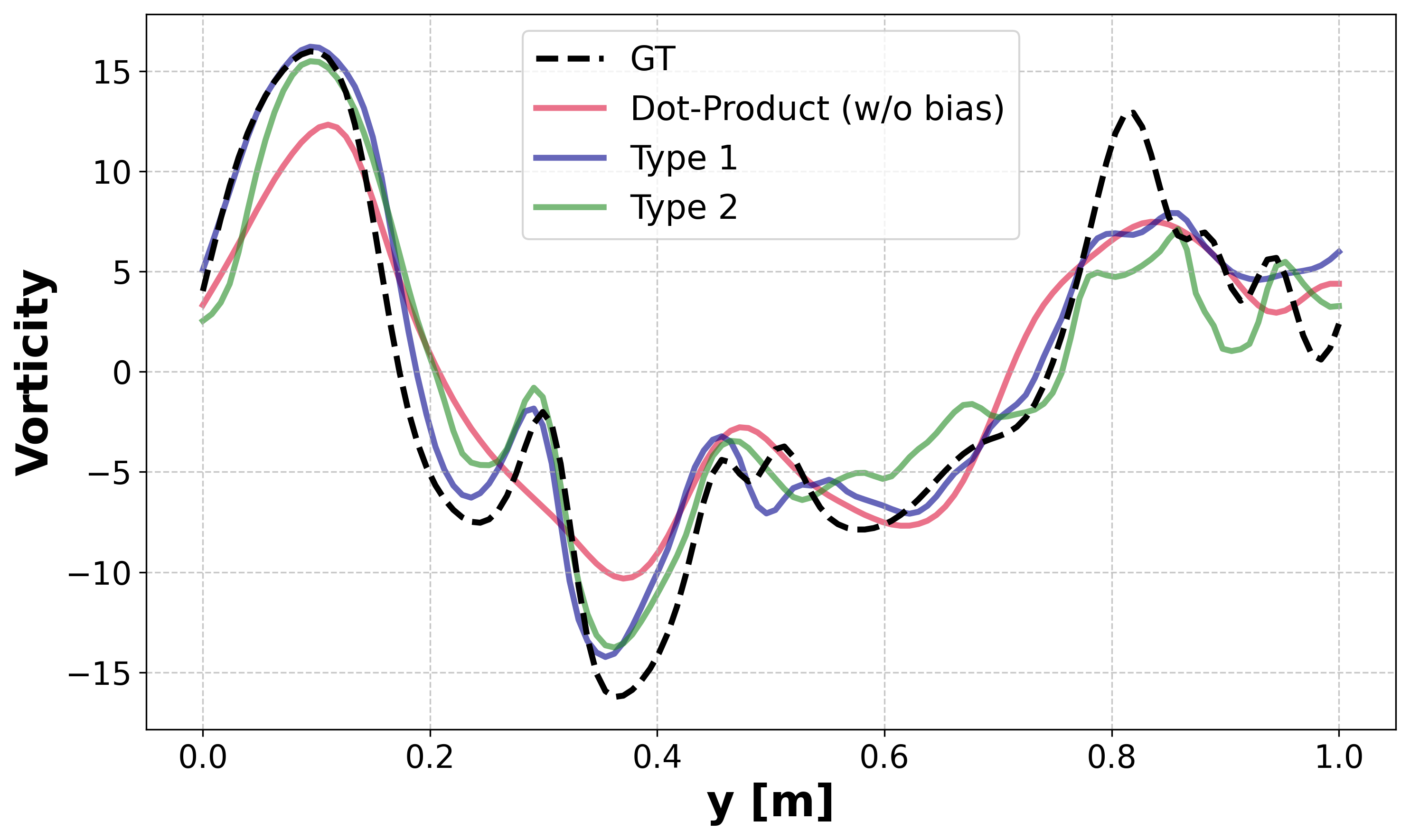}
        \caption{$t=0.75s$}
        \label{fig:merge_Sampling_lineplot2}
    \end{subfigure}
    \caption{Comparison of vorticity predictions along $x=0.5m$ between different merge strategies. The plots show vorticity values versus y-coordinate for two different time steps, comparing predictions from dot-product operation (without bias) and merge network approaches (Type 1 and 2) against ground truth (GT). Both merge network types demonstrate superior ability in capturing local flow features and fluctuations compared to the basic dot-product operation.}
    \label{fig:merge_Sampling_lineplot}
\end{figure}

To better understand the qualitative differences between these merge strategies, we visualize the predicted vorticity profiles along a vertical line at $x=0.5m$ for different time steps in Figure~\ref{fig:merge_Sampling_lineplot}. These line plots show how vorticity values vary along the y-direction from $0$ to $1m$, providing insight into each model's ability to capture flow field features at different spatial locations. The visualization clearly demonstrates the limitations of the dot-product operation and the advantages of incorporating merge networks. At both time steps (Figure~\ref{fig:merge_Sampling_lineplot1} and Figure~\ref{fig:merge_Sampling_lineplot2}), the dot-product approach (without bias, which outperformed the with-bias variant) exhibits significant difficulties capturing local flow features, producing overly smoothed predictions that miss important vorticity fluctuations across the domain. In contrast, both Type 1 and Type 2 merge networks successfully reproduce the complex flow patterns, following the ground truth's local peaks and valleys with notably higher fidelity.

\textbf{\textit{Note on overfitting issues in over-parameterized DeepONet}}: To obtain the architectures summarized in Table~\ref{tab:merge_comparison}, we conducted manual hyperparameter tuning across a wide range of model depths, hidden layer sizes, and merge configurations with varying parameter numbers. Despite testing deeper and wider networks with significantly larger parameter counts, we consistently observed that such models led to worse generalization performance---even as they achieved lower training losses. This suggests a critical insight: DeepONet tends to overfit easily when over-parameterized, indicating that its expressive power must be tightly controlled. The final architectures, each with approximately 300k parameters, were selected based on their superior balance between training efficiency and test accuracy.

\subsection{Effects of Spatial Point Subsampling and Application of Automatic Mixed Precision}\label{sec:pre_sampling}

Building upon the optimal merge network configuration (Type 1) identified in Section~\ref{sec:pre_merge}, we investigate the impact of spatial point subsampling and automatic mixed precision (AMP) training on model performance. Spatial point subsampling refers to our approach of selecting only a subset of available spatial locations from the full 128×128 grid for training, leveraging DeepONet's inherent capability to operate on arbitrary spatial points without requiring fixed grid structures. We explore three subsampling strategies: 16×16, 32×32, and 64×64 randomly selected points from the original 128×128 grid. The optimal choice of this subsampling ratio (e.g., the number of points like 16×16 or 32×32) can be highly problem-dependent, influenced by factors such as the physical characteristics of the investigated flow field, the scale of its primary dynamics, and overall data complexity. Therefore, for new applications, we recommend conducting a preliminary parametric study, similar to the exploration presented in Table~\ref{tab:sampling_comparison}, to determine a suitable subsampling level that effectively balances predictive accuracy with computational training efficiency.

AMP training refers to the strategic use of different numerical precisions (e.g., float32, float16) during model training to reduce memory usage and computational overhead without compromising model accuracy---a particularly relevant approach for DeepONet due to its intensive computational requirements in processing high-dimensional flow field data (for the details of AMP, please refer to Appendix~\ref{app:AMP}). For each subsampling configuration, we conduct experiments both with and without AMP training to systematically evaluate their combined effects on computational efficiency and prediction accuracy. All other training settings remain consistent with Section~\ref{sec:pre_merge}.

The results shown in Table~\ref{tab:sampling_comparison} reveal important insights about DeepONet's training efficiency through the combination of sampling strategies and AMP training. First, examining the impact of spatial point sampling, we observe that smaller sampling sizes can maintain prediction accuracy while significantly reducing computational costs. With AMP enabled, the 16×16 sampling strategy achieves comparable accuracy (test error: 17.419) to the baseline that uses all available points (test error: 17.079), while dramatically reducing the training time from 8.672 to 0.350 hours---a 96\% reduction. Similar trends are observed in cases without AMP, where the 16×16 sampling rather achieves a test error of 16.876 compared to 18.001 for full sampling due to its effective sampling nature compared to naive all-point-sampling approach.

\begin{table}[htb!]
\centering
\setlength{\abovecaptionskip}{10pt} % Spacing
\renewcommand{\arraystretch}{1.1}
\caption{Comparison of different spatial subsampling strategies with and without AMP training: Type 1 merge network is applied.}
\label{tab:sampling_comparison}
\begin{tabular*}{0.7\columnwidth}{@{\extracolsep{\fill}}c c c c}
\hline
AMP & Number of Sampled Points & Training Hour & Test Error \\ \hline
\multirow{4}{*}{\textbf{ON}} & \textbf{16×16} & 0.350 & 17.419 \\ \cline{2-4}
 & 32×32 & 0.785 & 18.184 \\ \cline{2-4}
 & 64×64 & 2.211 & 17.438 \\ \cline{2-4}
 & 128×128 & 8.672 & 17.079 \\ \hline
\multirow{4}{*}{OFF} & 16×16 & 0.377 & 16.876 \\ \cline{2-4}
 & 32×32 & 1.038 & 16.989 \\ \cline{2-4}
 & 64×64 & 3.458 & 17.862 \\ \cline{2-4}
 & 128×128 & 13.648 & 18.001 \\ \hline
\end{tabular*}
\end{table}

The comparison between AMP and non-AMP training reveals interesting trade-offs in computational efficiency and prediction accuracy across different sampling sizes. While AMP-OFF training with 16×16 sampling achieves slightly better accuracy (16.876 versus 17.419 with AMP-ON) and requires slightly longer training time (0.377 versus 0.350 hours), we opt for the AMP-ON configuration with 16×16 sampling in subsequent experiments. This choice is motivated by two practical considerations: (1) our comprehensive study involves numerous repeated experiments to account for training variability, where even small reductions in training time yield substantial cumulative savings; and (2) the computational advantage of AMP scales with sampling resolution---at full 128×128 sampling, AMP reduces training time by 36.5\%. This scalable efficiency demonstrates the practical value of AMP and underscores the importance of systematically exploring its effectiveness in various DeepONet settings to be investigated in later sections. Accordingly, the AMP-ON configuration with 16×16 sampling in Table~\ref{tab:sampling_comparison} is judged to represent the reasonable compromise between accuracy and computational cost in this study.

\begin{figure}[htb!]
\centering
\includegraphics[width=0.6\textwidth]{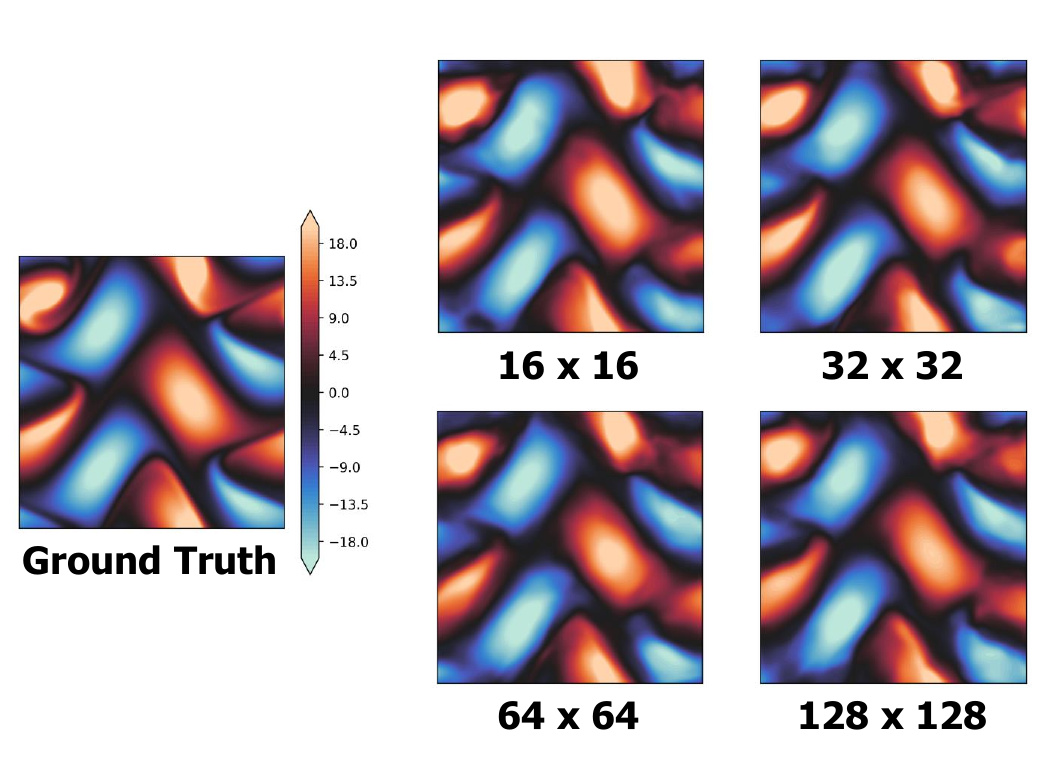}
\caption{To visually investigate the effects of number of subsampled points, 4 cases with AMP in Table~\ref{tab:sampling_comparison} are visualized: snapshots at $t=0.25s$. The comparison demonstrates that even with significantly reduced sampling (16×16), the model captures essential flow patterns similar to those observed in higher resolution samplings (32×32, 64×64) and the full dataset (128×128).}
\label{fig:diff_sampling_number}
\end{figure}

In Figure~\ref{fig:diff_sampling_number}, we can observe the effects of different subsampling resolutions on flow field prediction quality given AMP. The comparison demonstrates that even with the coarsest sampling resolution (16×16), the DeepONet successfully captures the essential vorticity patterns and intensity distributions present in the flow field at $t = 0.25s$. As expected, the visual differences between sampling resolutions are minimal, with all configurations reproducing the key flow structures observed in the ground truth. This visual evidence supports our quantitative findings in Table~\ref{tab:sampling_comparison}, which indicate that the 16×16 sampling strategy achieves comparable prediction accuracy (MSE of 17.419) to the full 128×128 sampling (MSE of 17.079), while dramatically reducing the training time from 8.672 to 0.350 hours. The consistent visual quality across different sampling resolutions highlights DeepONet's point-based prediction flexibility, which allows for efficient data utilization without substantial degradation in prediction performance.

The findings in this section highlight a fundamental advantage of DeepONet over grid-based architectures like CNNs or GNNs: its point-based prediction capability enables remarkable flexibility in data sampling, free from the constraints of fixed grid structures that limit traditional architectures and often require storing complete grid information in memory. This sampling flexibility, when combined with computational optimization techniques like AMP training, provides an effective strategy for enhancing DeepONet's training efficiency while significantly reducing memory requirements compared to grid-based methods. 

\subsection{Effects of Positional Encoding}\label{sec:pre_pe}

Building upon the optimal merge network configuration and sampling strategy identified in previous sections (merge Type 1 with 16×16 subsampling with AMP), we investigate the impact of positional encoding (PE) on DeepONet's prediction accuracy for spatio-temporal flow fields. While PE is a highly effective and well-established technique for coordinate-based networks in computer vision \cite{tancik2020fourier, mildenhall2021nerf, tewari2022advances, zheng2021rethinking, benbarka2022seeing, damodaran2023improved}, its application has predominantly focused on spatial coordinates. Consequently, clear guidelines for their optimal application to complex spatio-temporal problems within operator learning frameworks like DeepONet are less established. A key motivation for this part of our study, therefore, is to systematically explore its potential effectiveness and specific behaviors when applied to spatio-temporal prediction context. This investigation aims to provide valuable insights and practical guidelines for DeepONet users working at the intersection of these domains (e.g., CFD and computer science). To this end, we systematically explore three distinct PE strategies: (1) \textbf{PE($t$)}: applying PE only to the temporal coordinate ($t$); (2) \textbf{PE($x,y$)}: applying PE only to the spatial coordinates ($x,y$); and (3) \textbf{PE($x,y,t$)}: applying PE to all spatio-temporal trunk network inputs ($x,y$, and $t$). For each strategy, we evaluate combinations of two key hyperparameters: the scale factor $\sigma$ controlling the frequency range, and the mapping size $m$ determining the dimension of encoded features. Besides $\sigma$ and $m$, all other training settings remain consistent with previous settings. Though sigma values are explored in a greedy manner in this paper, please note that the selection of optimal $\sigma$ can be aided by the prior knowledge on the dataset: for example, if dominant temporal periods in the flow dynamics are known, $\sigma$ could be tuned to these frequencies to better capture temporal patterns. However, in more general cases where such prior knowledge is unavailable, a preliminary parametric study, analogous to our investigation summarized in Table~\ref{tab:pe_comparison}, is advisable.

\begin{table}[htb!]
\centering
\setlength{\abovecaptionskip}{10pt} % Spacing
\renewcommand{\arraystretch}{1.1}
\caption{Effects of positional encoding strategies and hyperparameters on DeepONet performance}
\label{tab:pe_comparison}
\begin{threeparttable}
% \begin{tabular*}{0.9\columnwidth}{@{\extracolsep{\fill}}l c c c c}
\begin{tabular}{l c c c c}
\hline
PE Strategy & Scale Factor ($\sigma$) & Mapping Size ($m$) & Training Hour & Test Error \\ \hline
\textbf{Without PE} & - & - & 0.350 & 17.419 \\ \hline
\multirow{9}{*}{\textbf{PE($t$)}}
& \multirow{3}{*}{1} & 16 & 0.355 & 18.271 \\ \cline{3-5}
& & 32 & 0.363 & 16.751 \\ \cline{3-5}
& & 64 & 0.363 & 17.331 \\ \cline{2-5}
& \multirow{3}{*}{5} & 16 & 0.355 & 17.797 \\ \cline{3-5}
& & 32 & 0.365 & 17.578 \\ \cline{3-5}
& & 64 & 0.359 & 18.065 \\ \cline{2-5}
& \multirow{3}{*}{\textbf{10}} & 16 & 0.358 & 16.619 \\ \cline{3-5}
& & \textbf{32} & 0.368 & \cellcolor{gray!30}\textbf{16.100} \\ \cline{3-5}
& & 64 & 0.357 & 17.421 \\ \hline
\multirow{9}{*}{\textbf{PE($x,y$)}}
& \multirow{3}{*}{1} & 16 & 0.398 & 21.255 \\ \cline{3-5}
& & 32 & 0.421 & 21.512 \\ \cline{3-5}
& & 64 & 0.405 & 20.658 \\ \cline{2-5}
& \multirow{3}{*}{5} & 16 & 0.460 & 25.196 \\ \cline{3-5}
& & 32 & 0.405 & 21.957 \\ \cline{3-5}
& & 64 & 0.422 & 20.978 \\ \cline{2-5}
& \multirow{3}{*}{10} & 16 & 0.410 & 117.874 \\ \cline{3-5}
& & 32 & 0.454 & 27.275 \\ \cline{3-5}
& & 64 & 0.405 & 22.090 \\ \hline
\multirow{9}{*}{\textbf{PE($x,y,t$)}}
& \multirow{3}{*}{1} & 16 & 0.356 & 21.709 \\ \cline{3-5}
& & 32 & 0.355 & 24.981 \\ \cline{3-5}
& & 64 & 0.352 & 24.197 \\ \cline{2-5}
& \multirow{3}{*}{5} & 16 & 0.354 & 43.161 \\ \cline{3-5}
& & 32 & 0.359 & 32.261 \\ \cline{3-5}
& & 64 & 0.351 & 29.300 \\ \cline{2-5}
& \multirow{3}{*}{10} & 16 & 0.358 & 81.884 \\ \cline{3-5}
& & 32 & 0.356 & 38.925 \\ \cline{3-5}
& & 64 & 0.306 & NaN\tnote{*} \\ \hline
\end{tabular}

\begin{tablenotes}\footnotesize
\item[*] Four out of five models diverged, indicating significant instability in this setting due to high-frequency embedding at spatial coordinates.
\end{tablenotes}

\end{threeparttable}
\end{table}

The comparison of different PE strategies in Table~\ref{tab:pe_comparison} reveals several important insights about DeepONet's behavior with coordinate encoding. First, it's noteworthy that all models demonstrate broadly similar training times (approximately 0.35-0.46 hours with NVIDIA A6000 GPU), indicating that the application of PE does not meaningfully impact computational efficiency. However, the results show a clear distinction between encoding strategies. The temporal-only strategy, PE($t$), improves model performance, with the best configuration ($\sigma=10$, $m=32$) achieving a test error of 16.100 compared to 17.419 without PE---a 7.57\% improvement. In contrast, applying PE to the spatial coordinates proved to be detrimental in our tests. The strategy of encoding all spatio-temporal inputs, PE($x,y,t$), led to deteriorated performance across all configurations. Also, experiments focusing only on spatial encoding, PE($x,y$), confirm this finding, as this strategy also consistently yielded higher errors than the baseline without PE (with a minimum test error of 20.658). This deterioration becomes particularly severe with higher scale factors for any strategy involving spatial encoding, culminating in training instability for PE($x,y,t$) at $\sigma=10$ and $m=64$, where four out of five training runs diverged.

The impact of hyperparameters reveals important considerations for PE implementation. The scale factor $\sigma$, which controls the frequency range of the encoding, shows strong influence on model performance. In PE($t$), higher $\sigma$ values generally lead to better performance, suggesting that temporal dynamics in our flow field benefit from higher-frequency encodings. Conversely, for strategies involving spatial coordinates (PE($x,y$) and PE($x,y,t$)), higher $\sigma$ values dramatically worsened performance, suggesting that high-frequency spatial embeddings destabilize the training process for this problem. This relationship between performance and $\sigma$ is highly dependent on the underlying characteristics of the spatio-temporal field, emphasizing the importance of careful hyperparameter tuning for specific applications. Furthermore, the mapping size $m$ demonstrates a non-monotonic relationship with model performance. For instance, in PE($t$), increasing $m$ from 32 to 64 consistently degrades performance across all scale factors, indicating that larger mapping sizes do not necessarily translate to better predictions. This suggests an optimal intermediate dimension for the encoded features that balances expressiveness with model complexity, with $m=32$ proving to be the optimal value in our case.

In conclusion, our results demonstrate that positional encoding, when properly configured and applied selectively to temporal coordinates, can enhance DeepONet's prediction accuracy without compromising computational efficiency. The 7.57\% improvement achieved with PE($t$) is a valuable addition for spatio-temporal prediction tasks, requiring no significant architectural or implementation overhead. However, the more critical insight from our systematic study is the finding that only temporal encoding is beneficial, whereas spatial encoding is detrimental for this problem. This stands in contrast to many computer vision applications where PE is successfully applied to spatial coordinates. Therefore, our investigation provides a valuable and non-obvious guideline: for spatio-temporal problems, one must carefully consider which coordinates to encode, as an indiscriminate application of PE can significantly impair model performance.

\section{Application of Multi-Fidelity DeepONet}\label{sec:MF} 

\subsection{Effects of Different Transfer Learning Approaches with Different Low-Fidelity Dataset}\label{sec:MF1} 

Building upon our optimized DeepONet architecture with temporal positional encoding ($\sigma=10$, $m=32$) in Section~\ref{sec:pre_pe}, we evaluate the following various MF-DeepONet architectures, all designed with the same model parameter counts to ensure fair comparison:

\begin{enumerate}
    \item \textbf{\citet{lu2022multifidelity} (conventional MF-DeepONet)}: while the original MF-DeepONet approach proposed by \citet{lu2022multifidelity} combines both residual learning and input augmentation within a coupled two-DeepONet framework, we strategically implement only the input augmentation aspect. We discard the coupled use of two DeepONet models and residual learning component due to their significant limitation: they fundamentally require identical query points across fidelity levels, which severely restricts the flexible usage of multi-fidelity datasets. Therefore, its implementation uses a sequential training architecture where, following the input augmentation strategy from \citet{lu2022multifidelity}, the output of the pre-trained LF model is used as an additional input for the HF model's trunk network. This architecture still necessitates that both networks execute in sequence during all training and inference operations, requiring much more computational time than our proposed framework.
    \item \textbf{Proposed}: as proposed in Section~\ref{sec:method_MFDeepONet2}, our MF-DeepONet employs fine-tuning where the pre-trained LF DeepONet's branch and trunk networks and frozen, while only the merge network parameters are updated during HF training. This approach preserves valuable LF representations while adapting specifically to HF features. This architecture only requires HF DeepONet during inference stage.
    \item \textbf{Full-tuning}: all network components (branch, trunk, and merge networks) are simultaneously trainable, potentially compromising the established low-fidelity knowledge while predominantly focusing on high-fidelity feature acquisition. This architecture only requires HF DeepONet during inference stage.
    \item \textbf{Linear probing}: takes a highly restrictive approach by freezing all layers except the final layer of the merge network. This architecture only requires HF DeepONet during inference stage.
\end{enumerate}

We systematically evaluate these approaches across different LF data configurations, varying both spatial resolution (16×16, 32×32, 64×64) and dataset size (50, 100, 200, 300 samples)---note that HF dataset size is fixed as 100. For fair comparison with the single-fidelity DeepONet which uses 800 training epochs, both LF and HF training phases in the multi-fidelity approaches are set to 400 epochs each, ensuring the total number of training epochs remains constant.

\begin{figure}[htb!]
    \centering
    % First (left) subplot
    \begin{subfigure}[b]{0.49\linewidth}
        \centering
        \includegraphics[width=\linewidth]{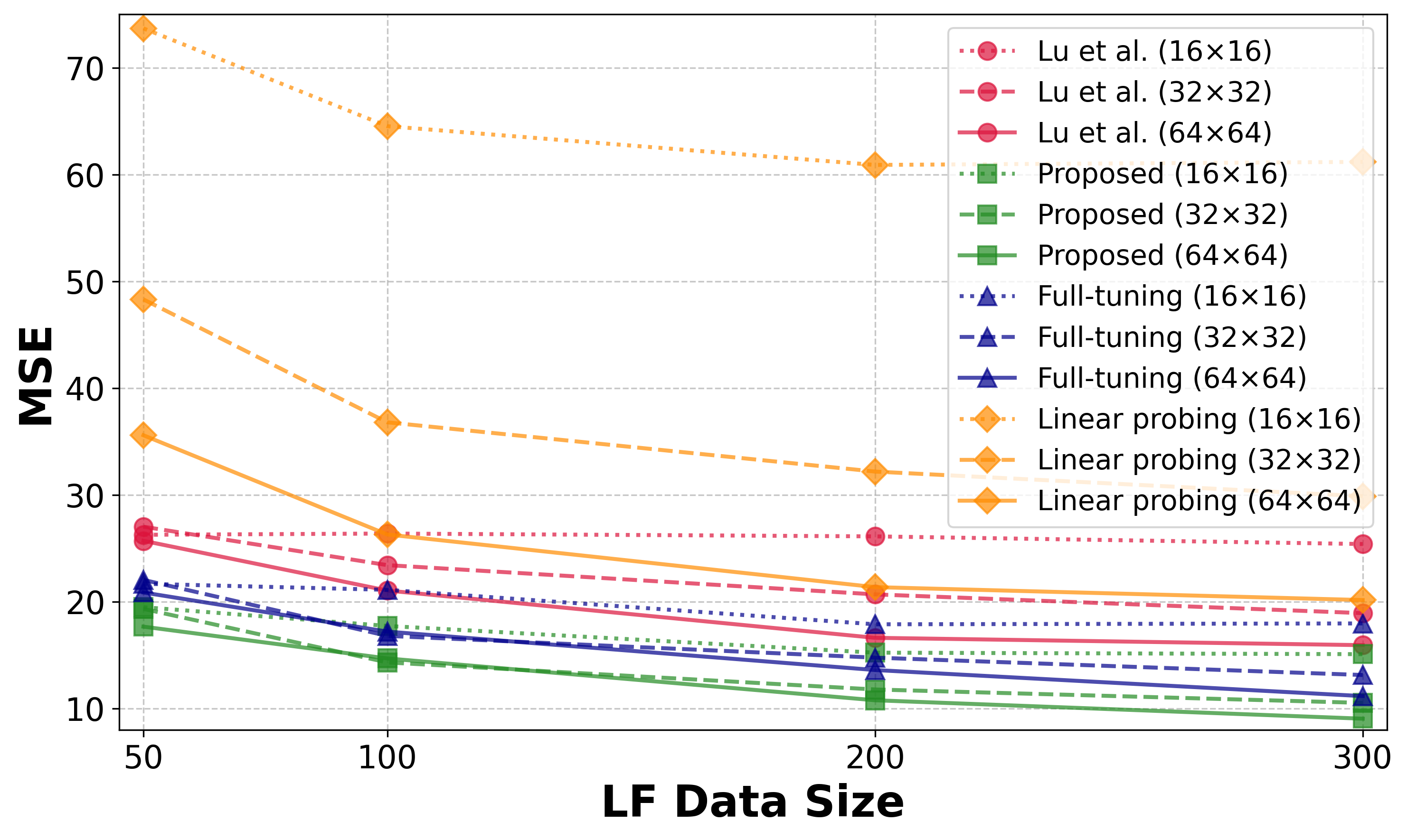}
        \caption{MSE}
        \label{fig:diff_tuning1}
    \end{subfigure}
    % Second (right) subplot
    \begin{subfigure}[b]{0.49\linewidth}
        \centering
        \includegraphics[width=\linewidth]{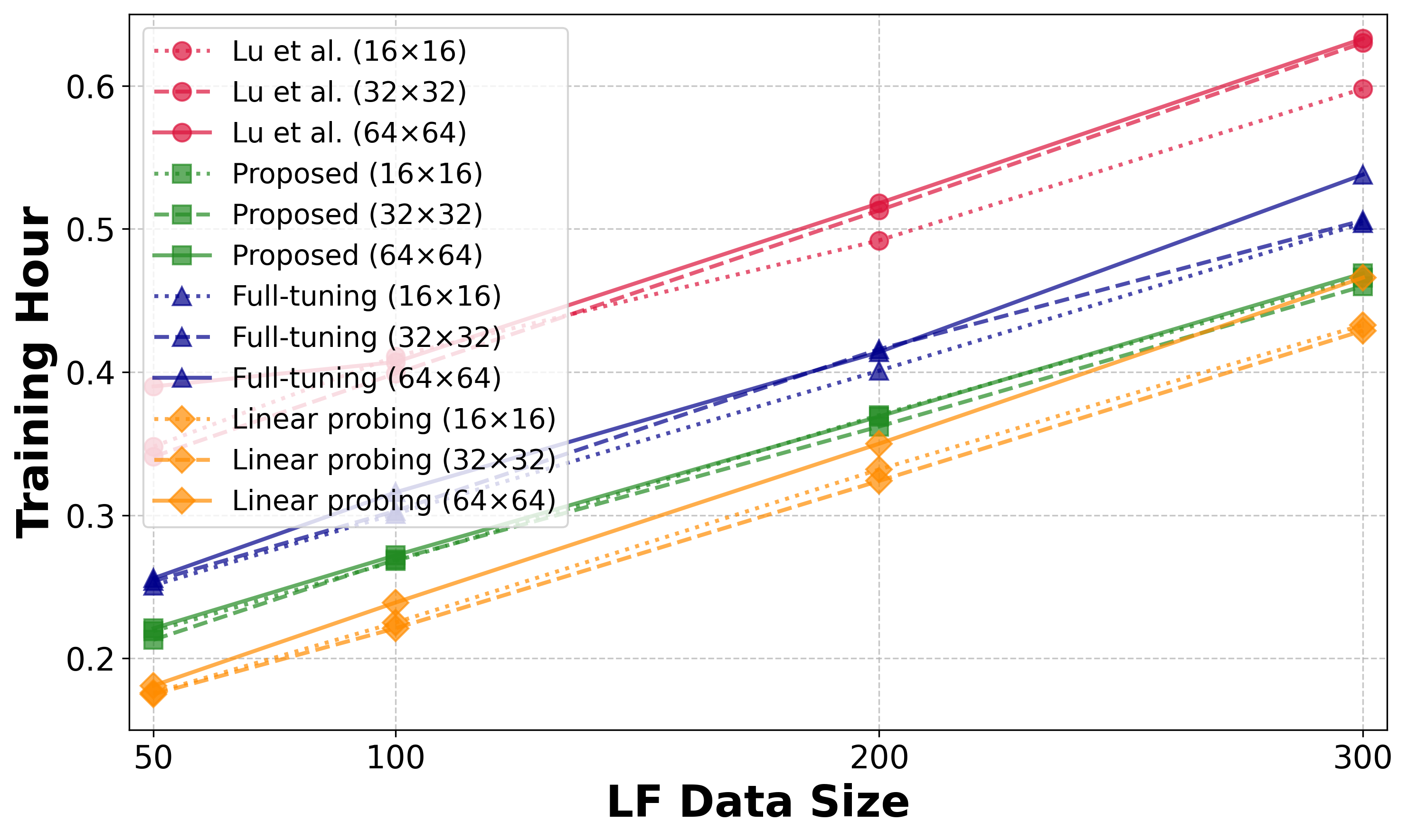}
        \caption{Training hour}
        \label{fig:diff_tuning2}
    \end{subfigure}

    \caption{Performance comparison of conventional DeepONet (\citet{lu2022multifidelity}), fine-tuning, full-tuning, and linear probing approaches for MF-DeepONet. Results show MSE and training time across different LF data sizes (50-300) and spatial resolutions (16×16: dotted, 32×32: dashed, 64×64: solid lines). This figure illustrates a visual representation of the numerical results presented in Table~\ref{tab:transfer_comparison}, enabling direct comparison of performance metrics across different model configurations.}
    \label{fig:diff_tuning}
\end{figure}

The results, presented in Table~\ref{tab:transfer_comparison} and also visualized as Figure~\ref{fig:diff_tuning} for easier comparison of MSE and training hour, reveal several insights about MF-DeepONet performance. First, to properly contextualize the benefits of a multi-fidelity approach, we establish the performance of single-fidelity models. Notably, training exclusively on low-fidelity data---even with a sufficient dataset size---proves insufficient for achieving competitive performance. For instance, single-fidelity models trained on 300 samples of 16×16, 32×32, and 64×64 data achieve MSE values of 70.154, 32.669, and 20.223, respectively. While the single-fidelity model trained on 100 high-fidelity samples performs better (MSE 16.100), all these single-fidelity cases are substantially outperformed by our best multi-fidelity model (MSE 9.057). This demonstrates that simply increasing the quantity of low-fidelity data cannot compensate for the fundamental need for high-fidelity information, highlighting the necessity of a multi-fidelity approach.

\begin{table}[htb!]
\centering
\setlength{\abovecaptionskip}{10pt} % Spacing
\renewcommand{\arraystretch}{1.}
\caption{Comparison of different MF-DeepONet architectures. Single-fidelity results are shown for reference (e.g., ``Single-fidelity (16×16, 300)'' indicates 300 training samples with 16×16 resolution).}
% The values highlighted in gray serve as baseline performance metrics for the analysis of time-derivative guided subsampling efficiency in Section~\ref{sec:TGS}.
\label{tab:transfer_comparison}
\begin{tabular}{ccccc}
\hline
Architecture & LF Data Resolution & LF Data Size & Training Hour & MSE \\
\hline
Single-fidelity (16×16, 300) & --- & --- & 0.476 & 70.154 \\
Single-fidelity (32×32, 300) & --- & --- & 1.257 & 32.669 \\
Single-fidelity (64×64, 300) & --- & --- & 1.272 & 20.223 \\
Single-fidelity (128×128, 100) & --- & --- & 0.368 & 16.100 \\
\hline
\multirow{12}{*}{\makecell{\citet{lu2022multifidelity}\\(conventional MF-DeepONet)}} 
& \multirow{4}{*}{16×16} & 50 & 0.348 & 26.273 \\
& & 100 & 0.411 & 26.388 \\
& & 200 & 0.492 & 26.123 \\
& & 300 & 0.598 & 25.407 \\
\cline{2-5}
& \multirow{4}{*}{32×32} & 50 & 0.341 & 27.029 \\
& & 100 & 0.399 & 23.432 \\
& & 200 & 0.513 & 20.694 \\
& & 300 & 0.630 & 18.936 \\
\cline{2-5}
& \multirow{4}{*}{64×64} & 50 & 0.390 & 25.710 \\
& & 100 & 0.407 & 21.053 \\
& & 200 & 0.518 & 16.628 \\
& & 300 & 0.633 & 15.940 \\
\hline
\multirow{12}{*}{\textbf{Proposed (fine-tuning)}} 
& \multirow{4}{*}{16×16} & 50 & 0.219 & 19.494 \\
& & 100 & 0.268 & 17.739 \\
& & 200 & 0.370 & 15.238 \\
& & 300 & 0.466 & 15.096 \\
\cline{2-5}
& \multirow{4}{*}{32×32} & 50 & 0.213 & 19.336 \\
& & 100 & 0.269 & 14.324 \\
& & 200 & 0.362 & 11.791 \\
& & 300 & 0.460 & \cellcolor{gray!30}10.536 \\
\cline{2-5}
& \multirow{4}{*}{\textbf{64×64}} & 50 & 0.221 & 17.673 \\
& & 100 & 0.272 & 14.698 \\
& & 200 & 0.369 & 10.782 \\
& & \textbf{300} & \textbf{0.469} & \cellcolor{gray!30}\textbf{9.057} \\
\hline
\multirow{12}{*}{Full-tuning}
& \multirow{4}{*}{16×16} & 50 & 0.251 & 21.687 \\
& & 100 & 0.301 & 21.119 \\
& & 200 & 0.401 & 17.893 \\
& & 300 & 0.504 & 17.974 \\
\cline{2-5}
& \multirow{4}{*}{32×32} & 50 & 0.254 & 22.050 \\
& & 100 & 0.303 & 16.807 \\
& & 200 & 0.416 & 14.772 \\
& & 300 & 0.506 & 13.146 \\
\cline{2-5}
& \multirow{4}{*}{64×64} & 50 & 0.256 & 20.870 \\
& & 100 & 0.316 & 17.208 \\
& & 200 & 0.414 & 13.612 \\
& & 300 & 0.538 & 11.171 \\
\hline
\multirow{12}{*}{Linear probing}
& \multirow{4}{*}{16×16} & 50 & 0.176 & 73.678 \\
& & 100 & 0.225 & 64.520 \\
& & 200 & 0.332 & 60.894 \\
& & 300 & 0.433 & 61.179 \\
\cline{2-5}
& \multirow{4}{*}{32×32} & 50 & 0.175 & 48.306 \\
& & 100 & 0.221 & 36.800 \\
& & 200 & 0.324 & 32.190 \\
& & 300 & 0.429 & 29.881 \\
\cline{2-5}
& \multirow{4}{*}{64×64} & 50 & 0.181 & 35.587 \\
& & 100 & 0.239 & 26.301 \\
& & 200 & 0.350 & 21.373 \\
& & 300 & 0.466 & 20.163 \\
\hline
\end{tabular}
\end{table}

With these baselines established, we proceed to compare the multi-fidelity methods. Our proposed fine-tuning approach consistently outperforms all other methods, including the input augmentation implementation of the conventional MF-DeepONet (\citet{lu2022multifidelity}). It shows limited effectiveness, with MSE values ranging from 25.407 to 26.388 for 16×16 resolution data, significantly underperforming even single-fidelity training with high-resolution data (MSE 16.100). Even at higher resolutions (64×64 LF data with 300 samples), this approach achieves an MSE of only 15.940, which is 76\% higher than our fine-tuning method's 9.057 MSE. While these inferior results from conventional MF-DeepONet may partly stem from our necessary omission of the residual learning component from the original \citet{lu2022multifidelity} framework, this design choice was deliberate and essential for our study's objectives. The residual learning approach, while potentially beneficial in certain contexts, fundamentally restricts DeepONet's inherent data flexibility by requiring identical sampling points across fidelity levels---a constraint that severely limits practical applicability in real engineering scenarios where sampling locations often differ between low and high-fidelity simulations (indeed, input function parameters, $var_1$ and $var_2$, are not nested across fidelity levels in this study). Additionally, the sequential architecture of \citet{lu2022multifidelity}'s approach---requiring both LF and HF DeepONet networks to be executed in sequence during inference---leads to consistently higher computational costs, with training times approximately 25-35\% longer than our proposed method across all configurations.

Second, fine-tuning consistently outperforms both full-tuning and linear probing across all configurations, demonstrating the effectiveness of our strategy to preserve learned low-fidelity latent representations by branch/trunk networks while adapting only the merge network to high-fidelity features. Particularly noteworthy is that even with the coarsest LF resolution (16×16), fine-tuning achieves better performance compared to single-high-fidelity training (MSE of 15.238 vs 16.100) when provided with sufficient LF data (200 samples). This improvement is unexpected given that the 16×16 resolution data exhibits remarkably different flow physics compared to higher resolutions, as evidenced in Figure~\ref{fig:data1}. While full-tuning provides more flexibility through complete network adaptation, this additional freedom may not be optimal, as it can lead to less efficient use of valuable low-fidelity feature representations learned during pre-training. The superior performance of fine-tuning suggests that the branch and trunk networks effectively capture essential input function parameter and spatio-temporal representations during low-fidelity training, which can be successfully leveraged for high-fidelity predictions through careful adaptation of the merge network alone.

\begin{figure}[htb!]
    \centering
    % First (top) subplot
    \begin{subfigure}[b]{0.95\linewidth}
        \centering
        \includegraphics[width=\linewidth]{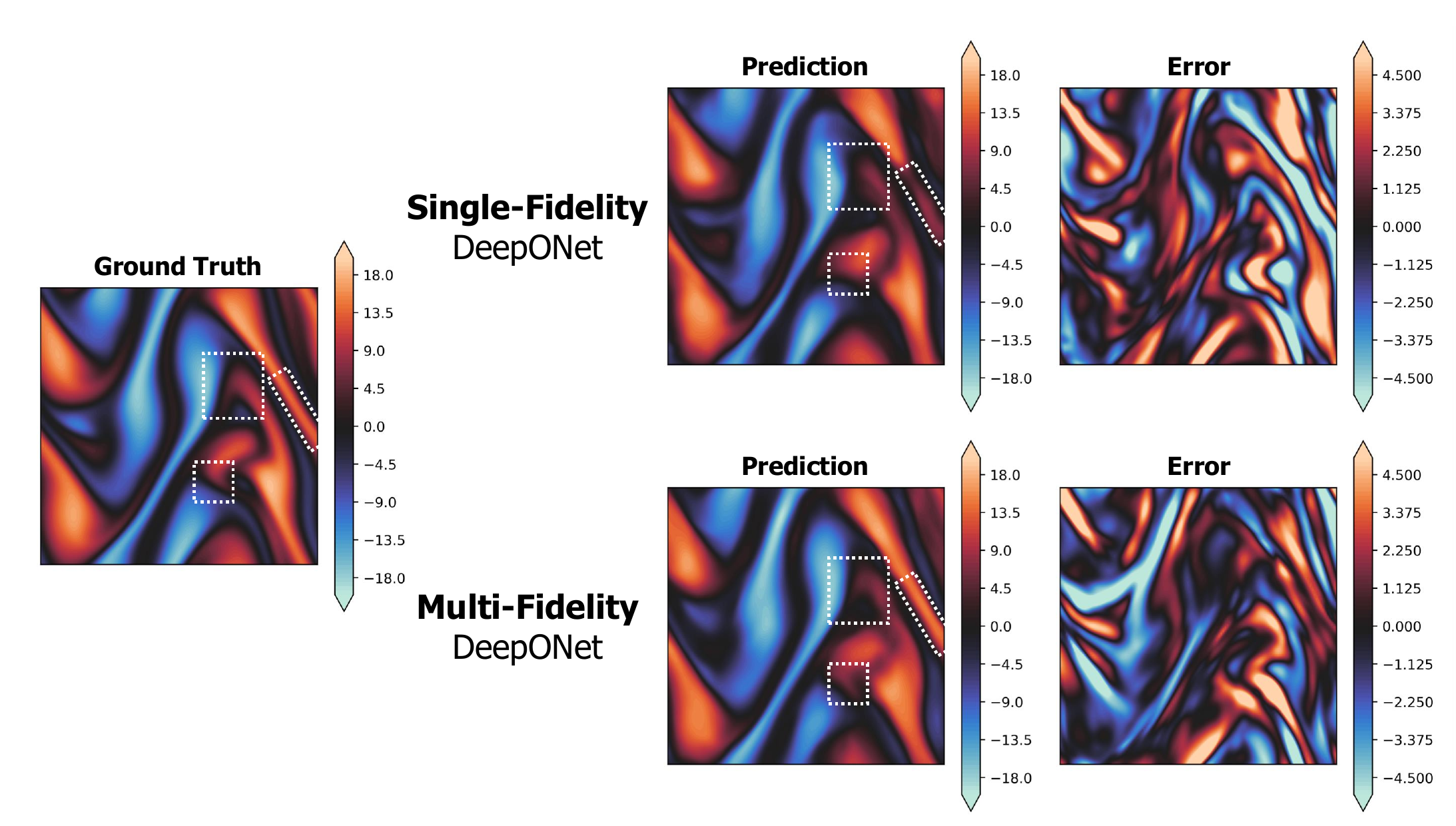}
        \caption{$t=0.5s$}
        \label{fig:SFvsMF1}
    \end{subfigure}
    \vspace{0.5em}

    % Second (bottom) subplot
    \begin{subfigure}[b]{0.95\linewidth}
        \centering
        \includegraphics[width=\linewidth]{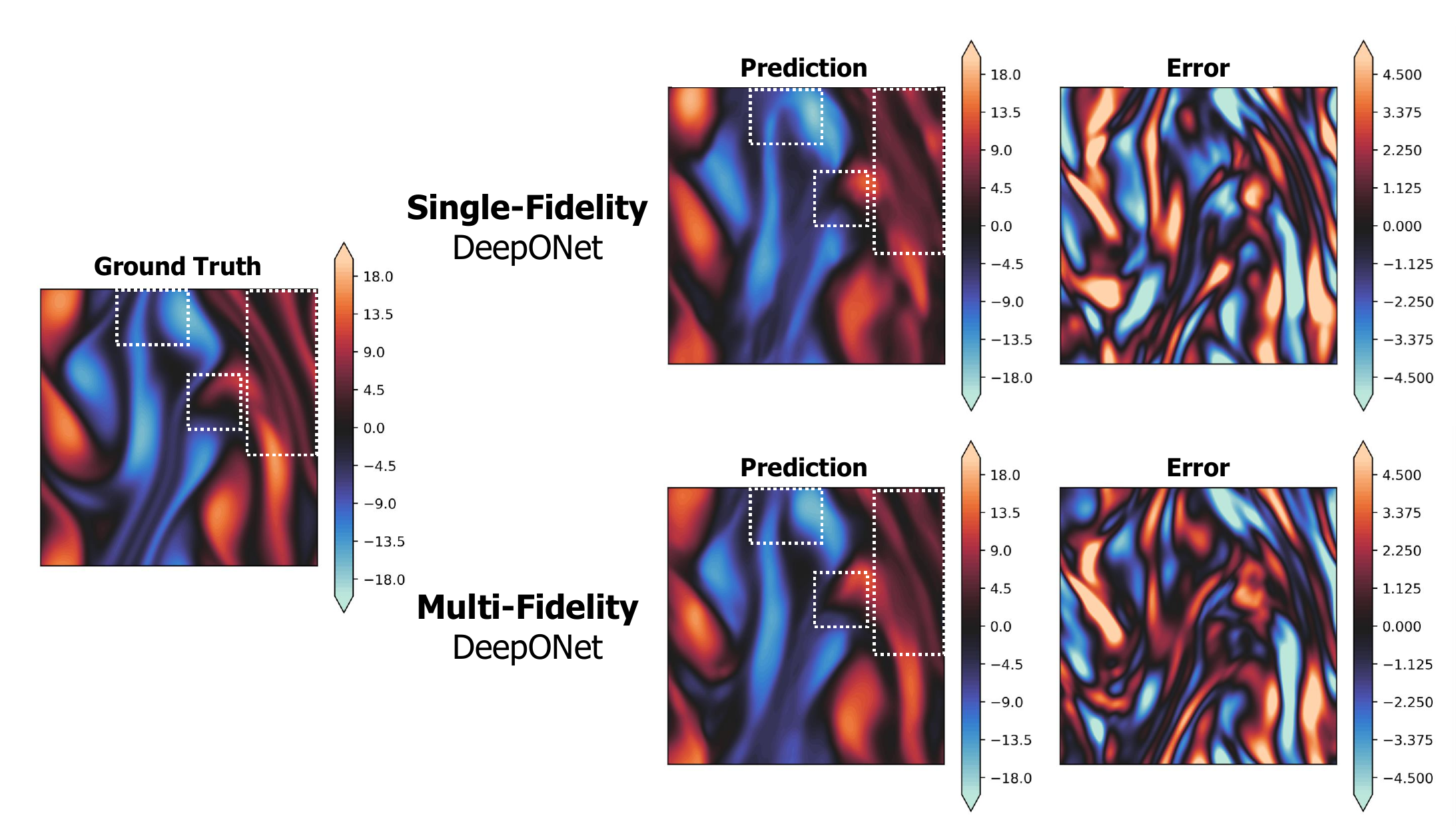}
        \caption{$t=0.75s$}
        \label{fig:SFvsMF2}
    \end{subfigure}

    \caption{Visual comparison of the predicted vorticity field at $t=0.5s$ and $t=0.75s$ between single-fidelity (128×128 resolution) DeepONet and multi-fidelity DeepONet with fine-tuning (with LF data resolution of 64×64 and data size of 300) within test scenario.}
    \label{fig:SFvsMF}
\end{figure}

In terms of LF dataset, as expected, increasing the resolution of LF data from 16×16 to 64×64 leads to consistent improvements in prediction accuracy across all transfer learning approaches. With 300 LF samples at 64×64 resolution, fine-tuning achieves an MSE of 9.057, representing a 43.7\% improvement over single-high-fidelity training (where MSE is 16.001). Figure~\ref{fig:SFvsMF} presents their comparative visualization of vorticity field predictions at two different time steps ($t=0.5s$ and $t=0.75s$). Our proposed MF-DeepONet with fine-tuning demonstrates notably superior prediction accuracy compared to the single-fidelity approach, particularly in regions highlighted by white dotted boxes. Within these regions, the single-fidelity model exhibits significant deviations from the ground truth, failing to capture the correct vorticity patterns and flow structures. In contrast, our proposed MF-DeepONet successfully reproduces the complex flow features, maintaining fidelity to the true flow dynamics across both temporal snapshots. 
% This trend again suggests that the benefits of preserving learned low-fidelity representations while selectively adapting the merge network are fundamental to effective multi-fidelity learning. Also, the configuration using 64×64 resolution with 200 LF samples emerges as a particularly practical alternative to single-fidelity training, as it achieves superior accuracy (MSE of 10.782) while maintaining comparable computational cost (training time of 0.369 hours versus 0.368 hours for single-fidelity). This trend confirms that higher-fidelity low-resolution data, 64×64, provides more valuable initialization for the network parameters, enabling better capture of the underlying flow physics.

\textbf{\textit{Note on how different physics across fidelities affects multi-fidelity performance}}: The results in Table~\ref{tab:transfer_comparison} reveal important insights about knowledge transfer between fidelity levels. Linear probing consistently performs poorly across all configurations, with MSE values substantially higher than both fine-tuning and single-fidelity approaches. This underperformance is scientifically significant---it confirms the substantial differences between low-fidelity and high-fidelity flow physics, as the frozen representations from low-fidelity models inadequately capture high-fidelity features without adaptation. Despite these fundamental differences, our fine-tuning approach demonstrates remarkable effectiveness---even when using coarse 16×16 resolution low-fidelity data with markedly different physical characteristics from ground truth physics, the model still outperforms training on HF data alone. Specifically, fine-tuning with 16×16 resolution low-fidelity data achieves MSE values of 15.238 (with 200 samples) and 15.096 (with 300 samples), representing improvements of 5.4\% and 6.2\% respectively over the single-fidelity baseline with 128×128 resolution (MSE: 16.100). This confirms that our selective adaptation approach effectively leverages useful information from low-fidelity simulations, even when they exhibit partially different physical characteristics compared to their high-fidelity counterparts. These results also highlight that while MF performance improves when the LF data better approximates HF physics (e.g., LF with 64×64 resolution), the proposed framework can still extract meaningful information from LF data of lower fidelity (e.g., LF with 16×16 resolution), provided the dataset is sufficiently large. This suggests that implicit correlations between LF and HF domains can still be harnessed through data-driven adaptation. Nevertheless, we still acknowledge that the effectiveness of the MF framework is sensitive to the overall quality of the LF data (Section~\ref{sec:limit-MF}), and this dependency should be carefully considered when applying the framework to problems with unreliable or weakly correlated LF sources.

\textbf{\textit{Note on computational efficiency of proposed DeepONet over HF simulations}}:
The total training time for all explored models of our proposed MF-DeepONet framework reported in Table~\ref{tab:transfer_comparison} is approximately 3.958 hours, with each model averaging around 0.33 hours of training time. Considering that generation of a single CFD simulation for one HF sample (128×128 resolution) takes about 0.1 hours, the computational effort to train one MF-DeepONet model is less than generating only four new HF simulations. This comparison highlights that the one-time computational cost of training our surrogate model represents an attractive alternative to generating additional HF simulation data, since the trained DeepONet model can subsequently produce flow field predictions virtually instantaneously, enabling rapid evaluations for downstream tasks such as design optimization, flow control, or uncertainty quantification. This inference efficiency can be especially beneficial in practical scenarios where repeated evaluations are necessary and HF simulations would otherwise become a computational bottleneck.

\subsection{Performance Analysis with Limited High-Fidelity Data}\label{sec:smallHF}

To further investigate the potential of our MF-DeepONet framework in scenarios where HF data is scarce, we conduct a systematic study using reduced HF dataset sizes. While our previous analysis in Section~\ref{sec:MF1} utilized 100 HF samples, here we examine the model's performance across a range of smaller HF dataset sizes: 20, 40, 60, and 80 samples. This investigation addresses a critical challenge in real-world engineering applications, where HF data is typically extremely limited. Based on our findings from Section~\ref{sec:MF1}, we focus on the more effective LF resolutions of 32×32 and 64×64, excluding the 16×16 resolution due to its relatively limited performance benefits. For each LF resolution, we explore two different LF dataset sizes (200 and 300 samples) to understand how the quantity of LF data affects the model's ability to compensate for limited HF data. This systematic approach will allow one to establish practical guidelines for efficiently allocating computational resources between LF and HF simulations in practical applications.

\begin{figure}[htb!]
    \centering
    % First (left) subplot
    \begin{subfigure}[b]{0.49\linewidth}
        \centering
        \includegraphics[width=\linewidth]{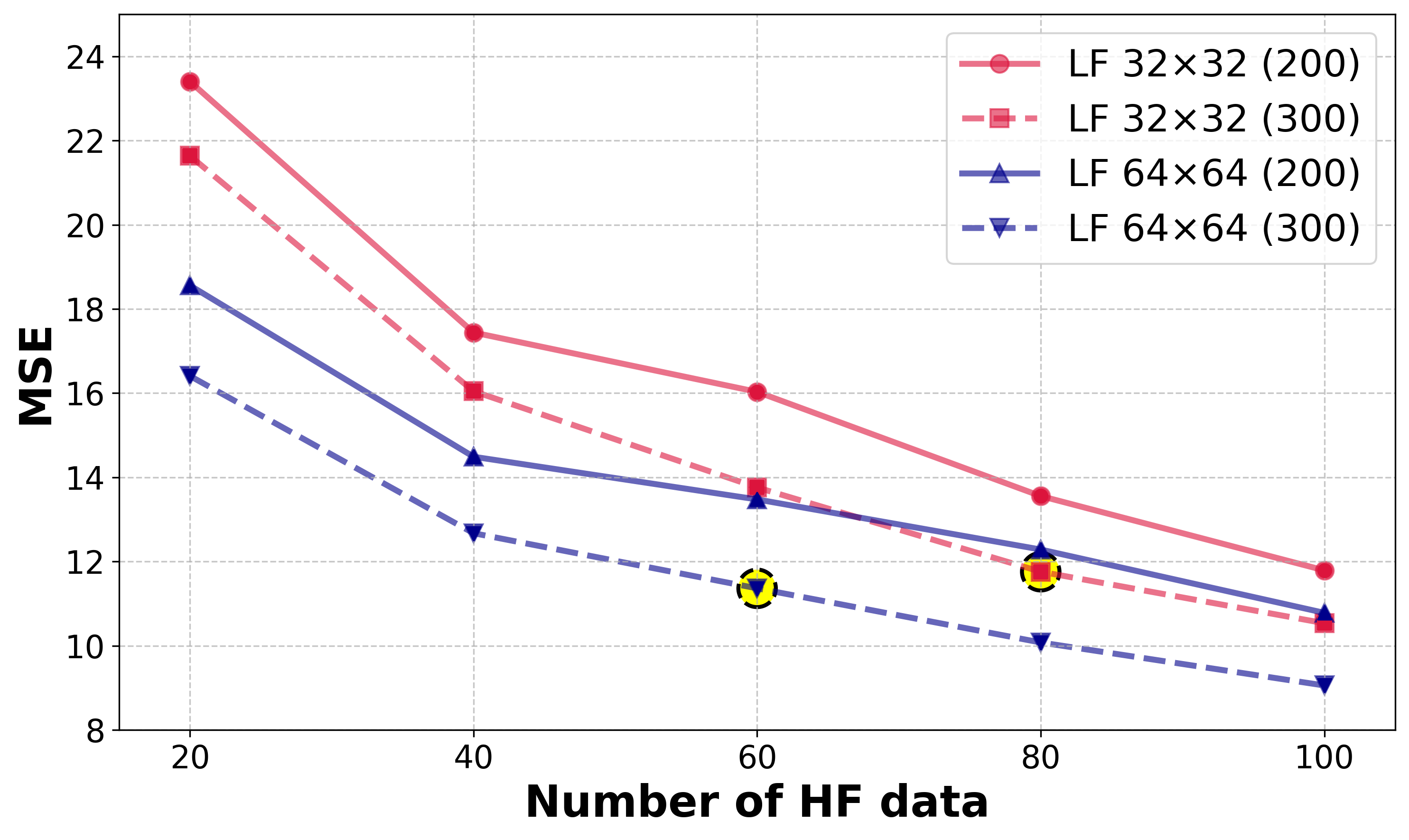}
        \caption{MSE}
        \label{fig:Limit_HF1}
    \end{subfigure}
    % Second (right) subplot
    \begin{subfigure}[b]{0.49\linewidth}
        \centering
        \includegraphics[width=\linewidth]{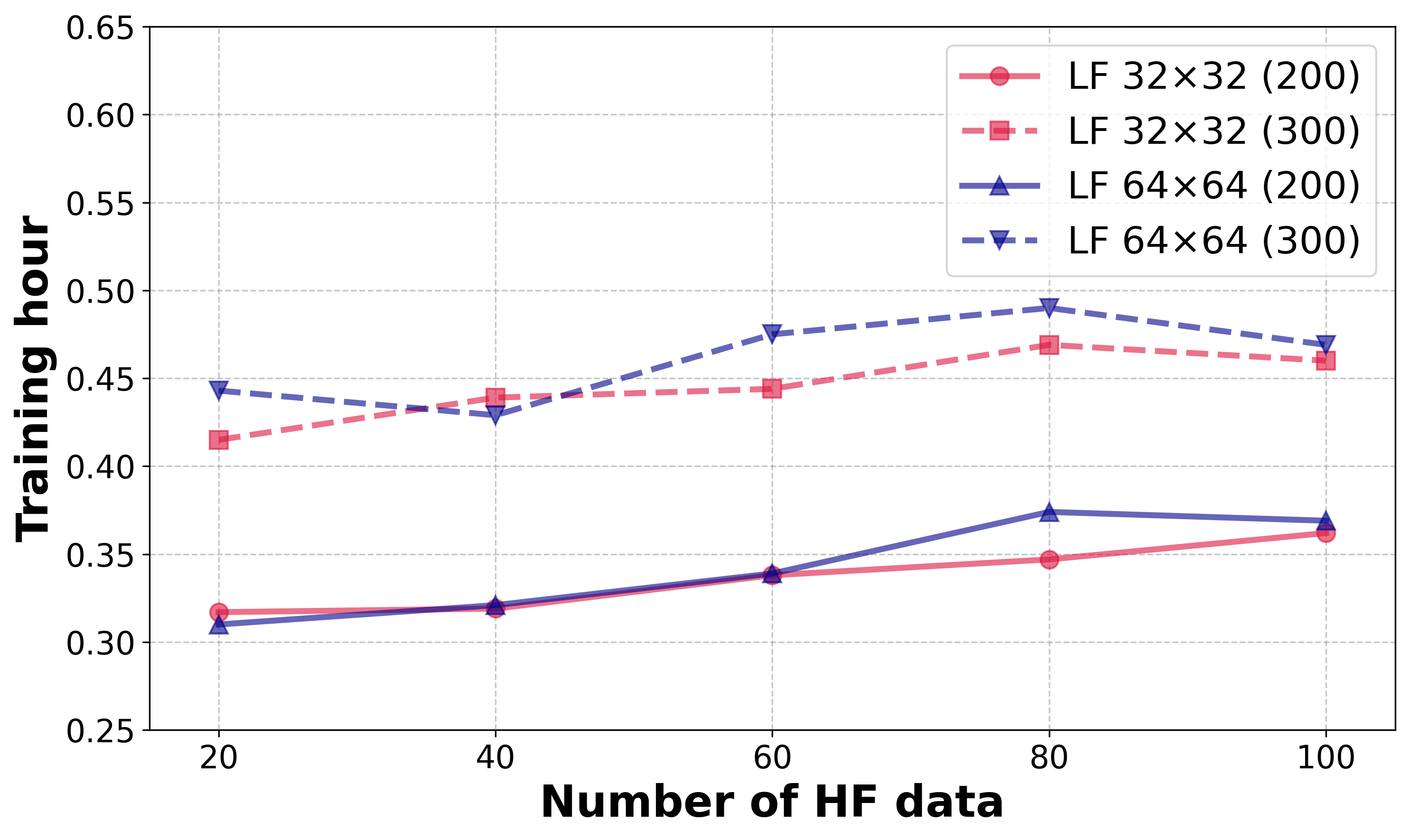}
        \caption{Training hour}
        \label{fig:Limit_HF2}
    \end{subfigure}

    \caption{MF-DeepONet performance with limited high-fidelity data: different resolutions of LF data with different LF data sizes are explored. The label ``LF 32×32 (200)'' represents a case where LF data is from a 32×32 resolution with a dataset size of 200---all other labels follow the same notation.}
    \label{fig:Limit_HF}
\end{figure}

The results shown in Figure~\ref{fig:Limit_HF} reveal several important insights about the MF-DeepONet's performance with limited HF data. First, examining the MSE trends (Figure~\ref{fig:Limit_HF1}), we observe a consistent pattern of improved prediction accuracy as the number of HF samples increases across all configurations. The 64×64 resolution configurations consistently outperform their 32×32 counterparts, with the 64×64 with 300 samples combination achieving the best performance across all HF dataset sizes. The training time analysis (Figure~\ref{fig:Limit_HF2}) reveals an expectable trade-off. Configurations with 300 LF samples consistently require longer training times compared to their 200 LF sample counterparts due to the longer training time required in LF deepONet. However, this increased computational cost appears justified by the improved accuracy, particularly for the 64×64 resolution cases.

% \begin{table}[htb!]
% \centering
% \caption{Multi-fidelity DeepONet performance with limited high-fidelity data: different resolutions of LF data with different LF data sizes are explored.}
% \label{tab:small_hf_comparison}
% \begin{tabular}{ccccc}
% \hline
% LF Data Resolution & LF Data Size & HF Data Size & Training Hour & MSE \\
% \hline
% \multirow{10}{*}{32×32} & \multirow{5}{*}{200} & 20 & 0.317 & 23.403 \\
% & & 40 & 0.319 & 17.441 \\
% & & 60 & 0.338 & 16.031 \\
% & & 80 & 0.347 & 13.556 \\
% & & 100 & 0.362 & 11.791 \\
% \cline{2-5}
% & \multirow{5}{*}{300} & 20 & 0.415 & 21.637 \\
% & & 40 & 0.439 & 16.047 \\
% & & 60 & 0.444 & 13.760 \\
% & & 80 & 0.469 & 11.759 \\
% & & 100 & 0.460 & 10.536 \\
% \hline
% \multirow{10}{*}{64×64} & \multirow{5}{*}{200} & 20 & 0.310 & 18.559 \\
% & & 40 & 0.321 & 14.490 \\
% & & 60 & 0.339 & 13.476 \\
% & & 80 & 0.374 & 12.286 \\
% & & 100 & 0.369 & 10.782 \\
% \cline{2-5}
% & \multirow{5}{*}{300} & 20 & 0.443 & 16.409 \\
% & & 40 & 0.429 & 12.674 \\
% & & 60 & 0.475 & 11.364 \\
% & & 80 & 0.490 & 10.078 \\
% & & 100 & 0.469 & 9.057 \\
% \hline
% \end{tabular}
% \end{table}

\section{Physics-Guided Subsampling for Enhancing Multi-Fidelity DeepONet}\label{sec:TGS}
\subsection{Effects of Physics-Guided Subsampling}\label{sec:sampling_results}

Based on our analysis in Section~\ref{sec:smallHF}, we will further investigate two configurations of our fine-tuned MF-DeepONet framework (highlighted in Figure~\ref{fig:Limit_HF1} as yellow circles with black dashed edges) that achieve an moderate balance between prediction accuracy and high-fidelity data efficiency:

\begin{enumerate}
\item \textbf{Case 1}: MF-DeepONet with 32×32 LF resolution (300 samples) and 80 HF samples
\item \textbf{Case 2}: MF-DeepONet with 64×64 LF resolution (300 samples) and 60 HF samples
\end{enumerate}

For these two configurations, we analyze the effectiveness of our physics-guided (specifically, time-derivative guided) sampling strategy in practice. This physics-guided approach, formalized in Algorithm~\ref{alg:time_deriv_sampling} in Section~\ref{sec:time_sampling}, uses a pre-trained LF DeepONet to identify regions with pronounced temporal dynamics. The purpose of this strategy is to make the training of the HF-DeepONet efficient by guiding the selection of a sparse but physically informative subset of spatial points from the full, already available HF simulation dataset, thereby enhancing data utilization efficiency during model training.

In our experiments, we first analyze the effectiveness of our proposed time-derivative guided subsampling. For a more comprehensive comparison, we also present the results from a conventional residual-based sampling strategy, which is a well-established technique in the domain of physics-informed neural networks \cite{lu2021deepxde}. This method operates by identifying and sampling points where the errors (residuals) between the LF model's predictions and the HF ground truth data are large. To ensure a fair comparison between these two guided approaches, we set the sampling ratio to $r=0.1$ for both, meaning that 10\% of the HF spatial points are strategically selected based on the respective criterion, while the remaining 90\% are sampled uniformly at random to maintain spatial coverage.

The results, detailed in Table~\ref{tab:dynamic_sampling}, demonstrate that while both methods significantly improve accuracy over the baseline, our proposed time-derivative subsampling is slightly superior in both accuracy and computational cost. Our physics-guided approach achieves MSE reductions of 10.02\% for Case 1 and 20.73\% for Case 2. In comparison, the residual-based method yields slightly smaller MSE reductions of 9.61\% and 19.74\% for the same cases. Furthermore, our approach is more computationally efficient, with an 8-11\% increase in training time overhead, compared to the more significant 18\% increase required for residual sampling. This additional overhead for both methods is primarily attributed to the calculations required to guide the sampling (i.e., automatic differentiation for temporal derivatives or residual computation). The greater improvement seen in the Case 2 configurations for both methods indicates that a higher-resolution LF data produces a more accurate LF DeepONet model, which in turn provides more reliable guidance.

\begin{table}[htb!]
\centering
\setlength{\abovecaptionskip}{10pt} % Spacing
\renewcommand{\arraystretch}{1.1} % Reverted to 1.1
\caption{Performance comparison of different sampling strategies: uniform (Before), proposed time-derivative guided (Time-Deriv. Sampling), and conventional residual-based (Residual).}
\label{tab:dynamic_sampling}
\begin{tabular}{ccccc}
\hline
Metric & Case & Before & Time-Deriv. Sampling & Residual Sampling \\
\hline
\multirow{2}{*}{MSE} & Case 1: LF 32×32 w/ 80 HF & 11.759 & 10.581 (-10.02\%) & 10.629 (-9.61\%) \\
\cline{2-5}
& Case 2: LF 64×64 w/ 60 HF & 11.364 & 9.008 (-20.73\%) & 9.121 (-19.74\%) \\
\hline
\multirow{2}{*}{Training hour} & Case 1: LF 32×32 w/ 80 HF & 0.469 & 0.505 (+7.68\%) & 0.555 (+18.34\%) \\
\cline{2-5}
& Case 2: LF 64×64 w/ 60 HF & 0.475 & 0.526 (+10.74\%) & 0.562 (+18.32\%) \\
\hline
\end{tabular}
\end{table}

\begin{figure}[htb!]
    \centering
    % First (left) subplot
    \begin{subfigure}[b]{0.49\linewidth}
        \centering
        \includegraphics[width=\linewidth]{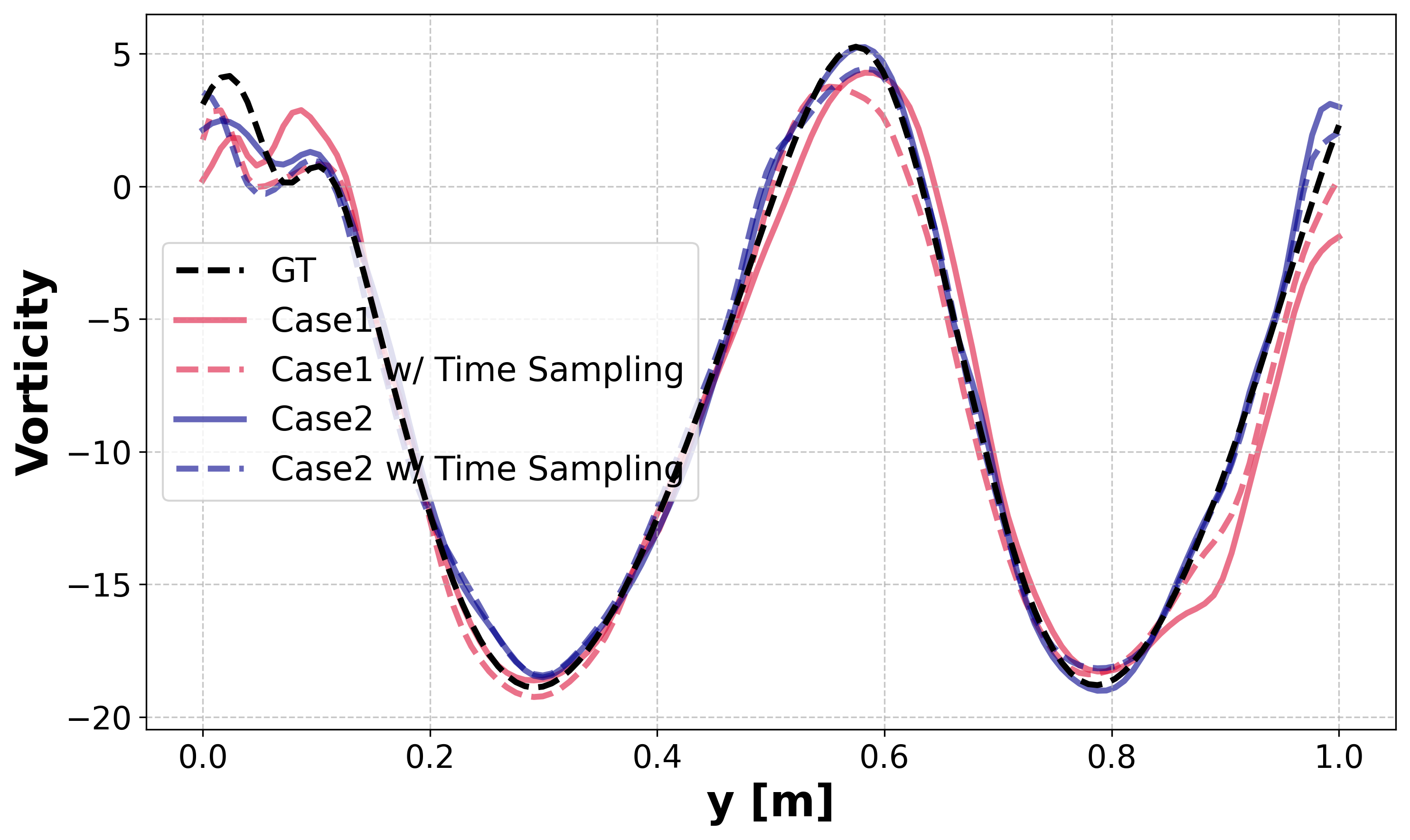}
        \caption{$t=0.25s$}
        \label{fig:Sampling_lineplot1}
    \end{subfigure}
    % Second (right) subplot
    \begin{subfigure}[b]{0.49\linewidth}
        \centering
        \includegraphics[width=\linewidth]{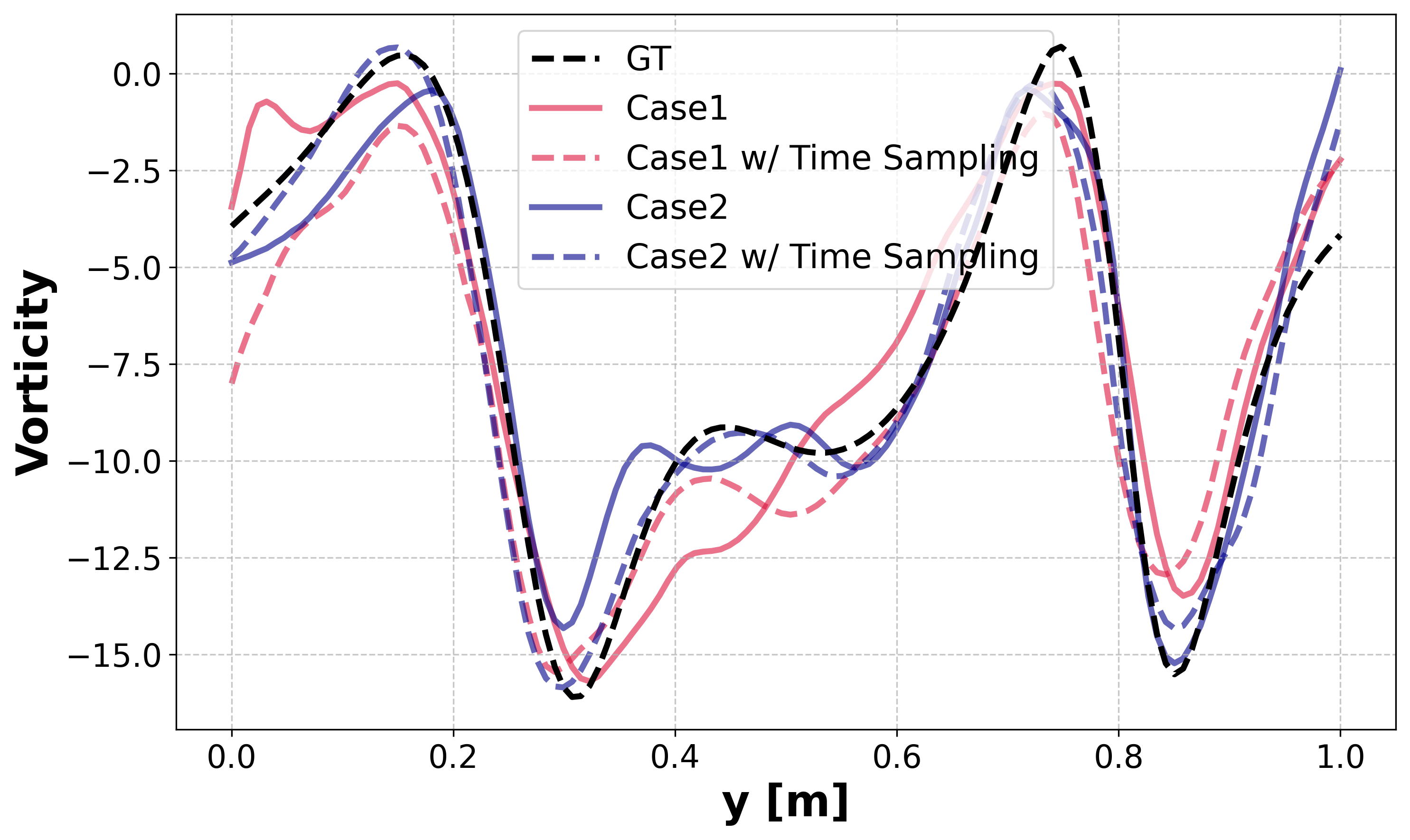}
        \caption{$t=0.5s$}
        \label{fig:Sampling_lineplot2}
    \end{subfigure}

    \caption{Comparison of vorticity profiles from ground truth (GT) and MF-DeepONet predictions with and without time-derivative guided subsampling. At early stages ($t=0.25s$), all models show comparable performance, but at later times ($t=0.5s$), models with time-derivative subsampling demonstrate superior accuracy in capturing complex flow dynamics. Case 1: 32×32 LF (300 samples) with 80 HF samples; Case 2: 64×64 LF (300 samples) with 60 HF samples.}
    \label{fig:Sampling_lineplot}
\end{figure}

The above quantitative improvements are further supported by the qualitative analysis shown in Figure~\ref{fig:Sampling_lineplot}, which compares vorticity profiles at different time steps. At $t=0.25s$ (Figure~\ref{fig:Sampling_lineplot1}), all models show similar prediction capabilities, closely following the ground truth profile. This comparable performance in the early stage is expected, as the flow field remains relatively simple and closer to the initial conditions. However, at $t=0.5s$ (Figure~\ref{fig:Sampling_lineplot2}), clear differences emerge in the models' ability to capture complex flow dynamics. This later time step exhibits notably more complicated flow development with pronounced temporal variations. Indeed, particularly in the middle region ($y=0.3m$ to $0.7m$), models trained with time-derivative subsampling (presented as dashed lines) more accurately reproduce the local fluctuations and subtle variations in the vorticity profile. This superior performance demonstrates that our approach automatically identifies and focuses computational resources on regions with significant temporal dynamics, enhancing prediction accuracy exactly where HF DeepONet should focus. In contrast, models without sampling tend to oversimplify these local features, capturing only the overall trend while missing important flow structure details. These results confirm that our physics-guided subsampling strategy effectively targets time-sensitive regions of the flow field ($t=0.5s$ than $t=0.25s$), showing particular advantage in predicting complex time-dependent dynamics.

\subsection{Reduced High-Fidelity Data Requirement through Physics-Guided Subsampling}\label{sec:small_HF}

The time-derivative guided subsampling strategy demonstrates remarkable effectiveness, achieving superior predictive accuracy compared to conventional uniform sampling when using the same amount of HF data (Section~\ref{sec:sampling_results}). Therefore, in this section, we investigate how this approach can reduce the required HF dataset size while maintaining comparable accuracy to models trained without time-derivative subsampling---quantifying the potential data efficiency gains enabled by our physics-guided subsampling method. For MF-DeepONet using 64×64 LF resolution data (Case 2), our sampling approach achieves an MSE of 9.008 with only 60 HF samples, surpassing the performance of models trained on the full 100 HF samples with uniform sampling (MSE 9.057, marked as gray in Table~\ref{tab:transfer_comparison}). This achievement is particularly significant as it represents both improved accuracy and a 40\% reduction in required HF data. Similarly, with 32×32 LF resolution data (Case 1), our time-based sampling achieves prediction accuracy (MSE 10.581) using only 80 HF samples, effectively matching the performance of training with the complete HF dataset and uniform sampling (MSE 10.536, marked as gray in Table~\ref{tab:transfer_comparison}).

\begin{figure}[htb!]
   \centering
   % First (top) subplot
   \begin{subfigure}[b]{0.95\linewidth}
       \centering
       \includegraphics[width=\linewidth]{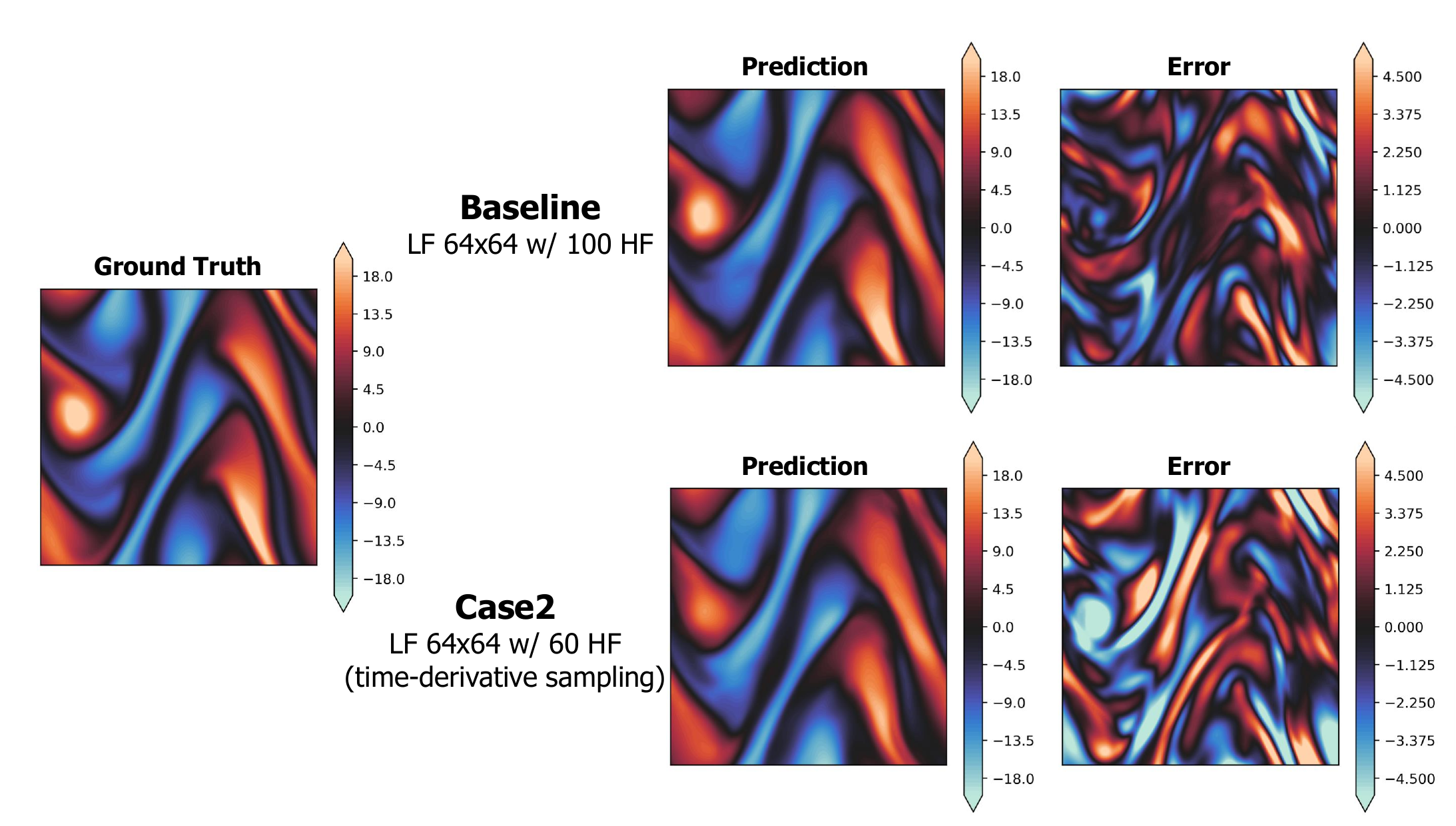}
       \caption{$t=0.5s$}
       \label{fig:ReducedHF1}
   \end{subfigure}
   \vspace{0.5em}
   % Second (bottom) subplot
   \begin{subfigure}[b]{0.95\linewidth}
       \centering
       \includegraphics[width=\linewidth]{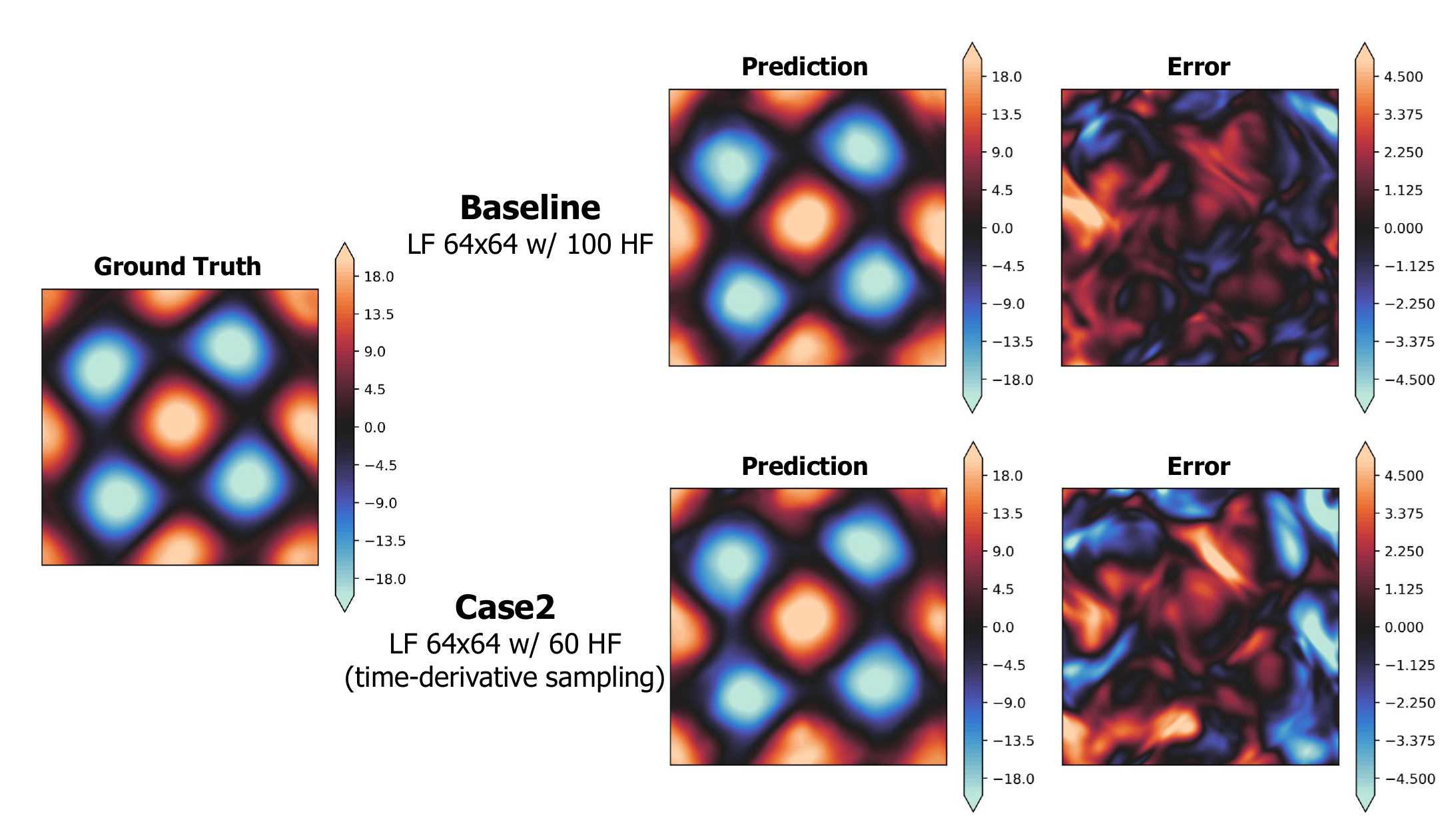}
       \caption{$t=0.75s$}
       \label{fig:ReducedHF2}
   \end{subfigure}
   \caption{Visual comparison of the predicted vorticity field at $t=0.5s$ and $t=0.75s$ between two models: 1) baseline uniform-sampling-based MF-DeepONet with LF 64×64 and 100 HF samples, and 2) MF-DeepONet with LF 64×64 and only 60 HF samples guided by time-derivative subsampling. Despite using 40\% fewer HF samples, the model with time-derivative subsampling maintains comparable prediction quality, with only minor differences in error patterns.}
   \label{fig:ReducedHF}
\end{figure}

Figure~\ref{fig:ReducedHF} provides a visual comparison between the uniform-sampling-based multi-fidelity model trained with the 100 HF dataset and the multi-fidelity model coupled with time-derivative guided subsampling approach using only 60 HF dataset. At both time steps ($t=0.5s$ and $t=0.75s$), we observe that both models produce visually comparable flow field predictions that closely match the ground truth. The error distributions demonstrate that, despite using significantly less HF data (only 60\%), the time-derivative subsampling approach maintains similar prediction quality to the uniform-sampling model. While there are subtle differences in the error patterns---particularly visible in certain regions at $t=0.75s$ where the proposed physics-guided subsampling approach shows slightly higher localized errors---these minor differences in error patterns are insignificant when considering the substantial 40\% reduction in HF data requirements achieved by our sampling approach. This visual evidence reinforces our quantitative findings that strategic selection of high-fidelity training points based on temporal dynamics enables efficient utilization of limited HF data with minimal degradation in prediction quality.

\begin{figure}[htb!]
\centering
\includegraphics[width=0.6\textwidth]{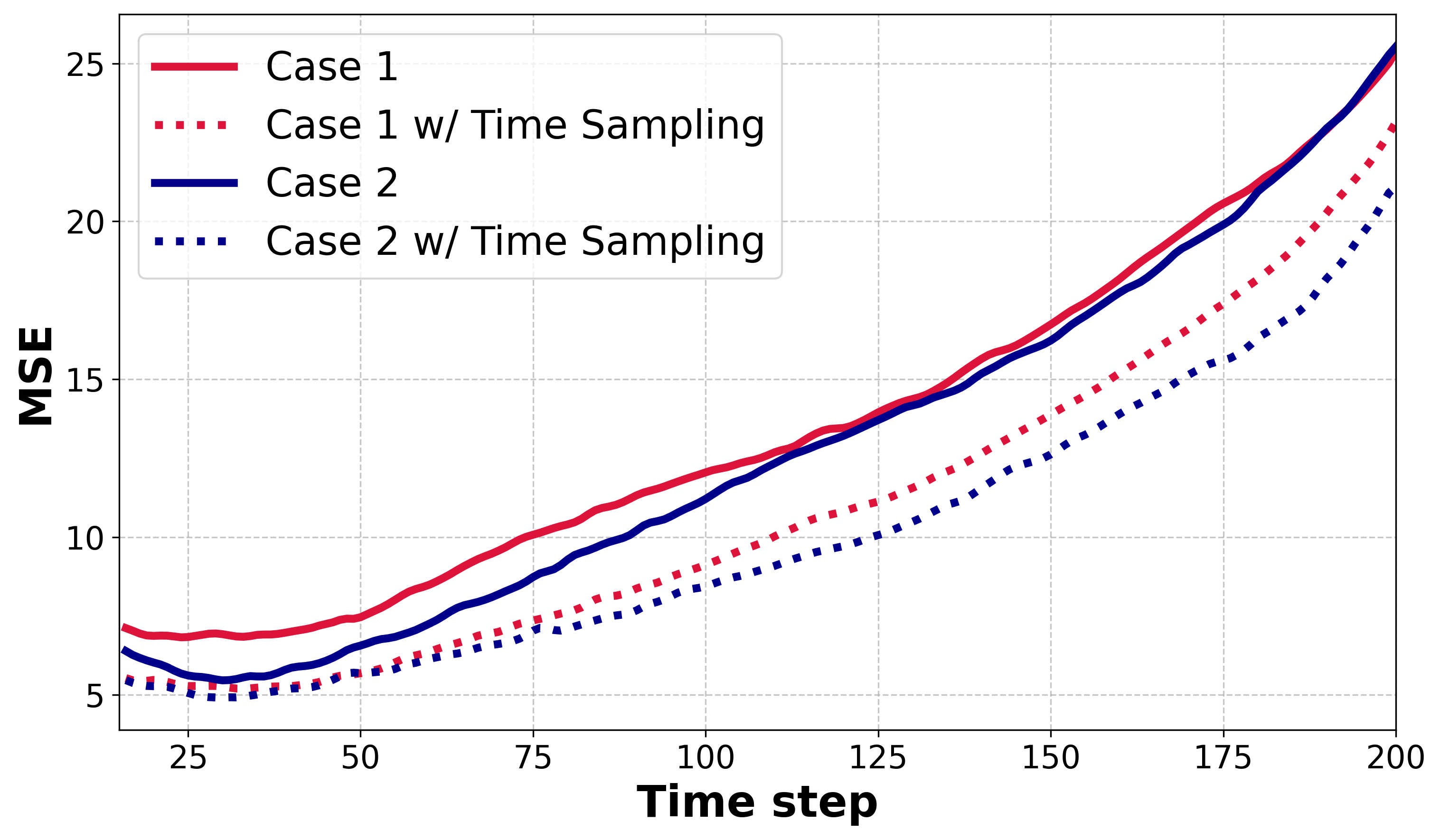}
\caption{Comparison of MSE behavior over time steps (200 steps stands for a second) for Case 1 and Case 2, with and without time-derivative guided subsampling. The results clearly demonstrate that the time-derivative guided subsampling effectively reduces error over time, thereby significantly improving prediction stability and accuracy.}
\label{fig:error_acc}
\end{figure}

Figure~\ref{fig:error_acc} presents an analysis of error behavior of the test data over time for Cases 1 and 2, comparing scenarios with and without time-derivative guided subsampling. It is clear from the plot that incorporating time-derivative guided subsampling mitigates error increase as temporal prediction progresses for both cases. This highlights that the physics-guided, time-derivative subsampling not only enhances accuracy at individual time steps but also substantially limits error propagation over extended prediction intervals. Such improved long-term predictive stability is particularly valuable for industrial applications where accuracy over prolonged simulation times is crucial.

\textbf{\textit{Note on the extensibility of the proposed physics-guided subsampling approach}}: While our sampling strategy leverages temporal derivatives as the key physical indicator, the MF-DeepONet framework itself is not confined to this specific sampling approach. Temporal derivatives serve as an effective physics-informed criterion for identifying critical regions in the examples studied here, capturing fundamental characteristics where rapid changes often indicate important physical transitions. However, the modular nature of our framework allows for integration with various alternative sampling techniques tailored to specific applications. For instance, future work could explore uncertainty-driven sampling for stochastic systems. This flexibility represents a clear advantage of our approach---the underlying MF-DeepONet architecture can be coupled with domain-specific sampling strategies while maintaining its core multi-fidelity benefits. This opens numerous avenues for researchers to develop specialized sampling techniques optimized for their particular physical systems, potentially yielding even greater efficiency gains beyond what we've demonstrated with temporal derivatives in this study.

\section{Further Validation with More Challenging Datasets}
\label{sec:newdata}

To further consolidate the robustness and superior performance of our proposed MF-DeepONet framework, particularly in scenarios demanding higher precision and involving more local flow dynamics, we introduce a new set of more challenging datasets. These datasets are designed to incorporate higher spatial resolutions and more complex initial conditions compared to those presented in Section~\ref{sec:data}. This section first validates our MF-DeepONet framework against conventional approaches using new datasets. Subsequently, we conduct a detailed investigation of our proposed time-derivative method, comparing it with a conventional residual-based approach to assess its efficacy under these more demanding conditions.

\subsection{Generation of Higher-Resolution and More Complex Flow Field Datasets}
\label{sec:newdata_generation}

The generation process for these new, more challenging datasets largely follows the methodology outlined in Section~\ref{sec:data}, utilizing the same pseudo-spectral solver for the incompressible Navier-Stokes equations in a two-dimensional periodic domain $[0,1m] \times [0,1m]$ with a Reynolds number of $Re=1,000$. However, two key aspects are modified:

\begin{enumerate}
    \item \textbf{More Complex Initial Conditions:} The parameters $var_1$ and $var_2$ defining the initial velocity field components (Eqn.~\ref{eq:ICs}) are now sampled from a uniform distribution in the range $\mathcal{U}[5,7]$ (previously $\mathcal{U}[1,3]$). This change introduces higher spatial frequencies into the initial state of the flow, leading to more complex and rapidly evolving flow structures.
    \item \textbf{Higher Spatial Resolutions:} The fidelity levels are increased accordingly as flow characteristics become complicated. For the HF dataset, we now generate 100 training samples and 50 test samples on a finer uniform grid of 256×256 resolution (previously 128×128). For the LF datasets, we generate 300 training samples for each of three upgraded resolutions: 32×32, 64×64, and 128×128 (previously 16×16, 32×32, and 64×64).
\end{enumerate}

These modifications therefore increase the computational cost of data generation. While HF 256×256 data requires approximately 1800 seconds per sample, the LF data requires 75, 120, and 330 seconds for 32×32, 64×64, and 128×128 resolutions, respectively. In total, generating the HF dataset (100 training samples) requires 50 hours, whereas generating each LF dataset (300 samples) requires 6.25, 10, and 27.5 hours, respectively.

\begin{figure}[htb!]
\centering
\includegraphics[width=0.65\textwidth]{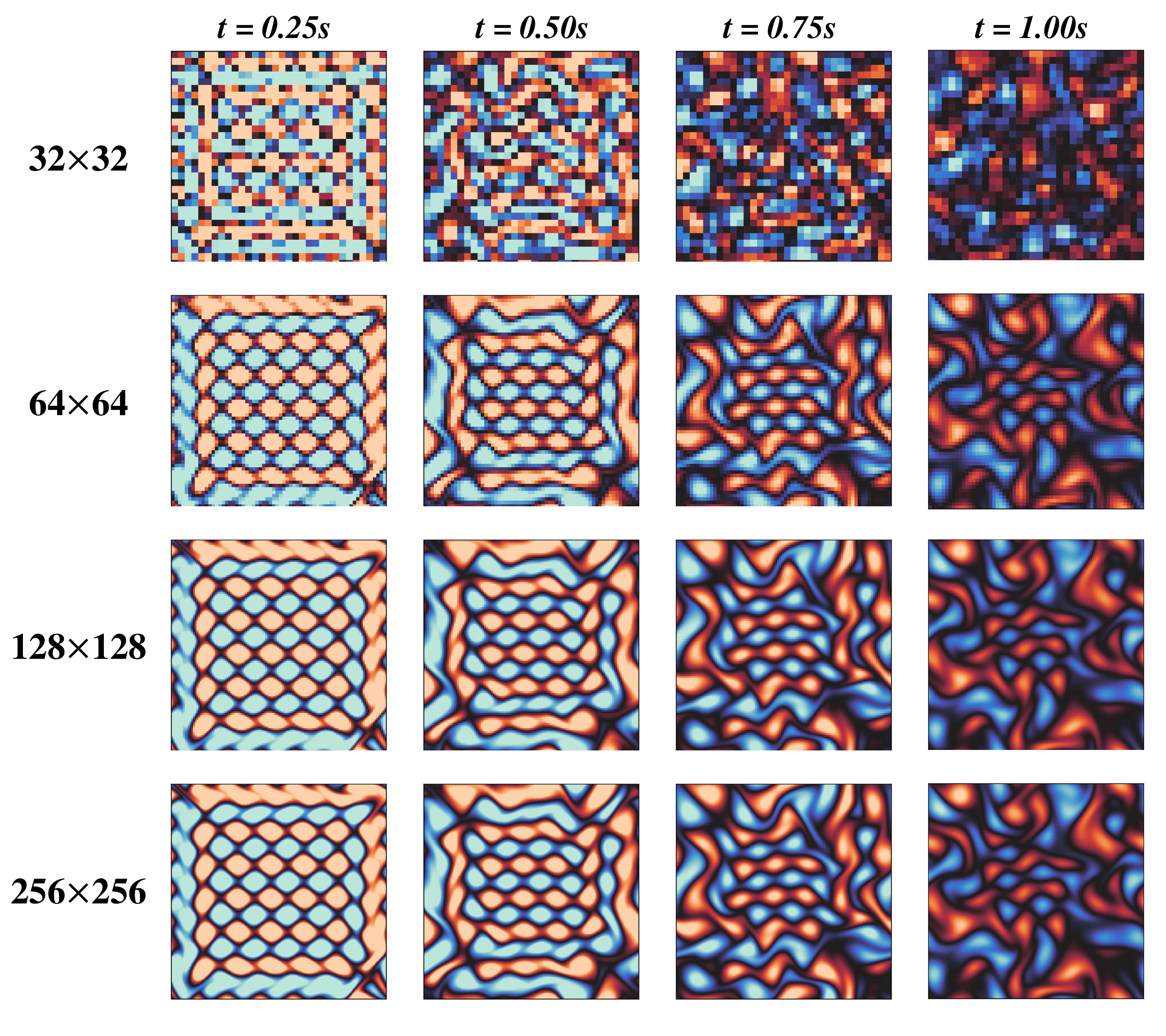}
\caption{Effects of grid resolution on vorticity field evolution with fixed parameters $var_1=6.4$ and $var_2=6.5$ (contours represent vorticity values).}
\label{fig:data2}
\end{figure}

The vorticity fields generated using these more challenging conditions are visualized across different resolutions in Figure~\ref{fig:data2}: this figure clearly illustrates the increased complexity compared to the initial dataset presented in Figure~\ref{fig:data1}. The higher values used for $var_1$ and $var_2$ in the initial conditions (Eqn.~\ref{eq:ICs}) result in a greater number of initial vorticity structures and consequently, a more intricate mixing process as the flow evolves. Similar to observations with the previous dataset, Figure~\ref{fig:data2} shows that the lowest LF resolution (32×32) struggles to represent the detailed flow structures visible in the HF (256×256) simulation, particularly at later time steps. In contrast, the intermediate LF resolutions (64×64 and 128×128) capture the essential macroscopic flow characteristics faithfully.

\subsection{Proposed MF-DeepONet Framework Performance}
\label{sec:MF_newdata}

Here, we validate the performance of our proposed MF-DeepONet framework against the conventional approach by \citet{lu2022multifidelity}. Since our initial results in Section~\ref{sec:MF1} conclusively showed that our fine-tuning strategy was superior to both full-tuning and linear probing, we focus this validation on our best-performing method (fine-tuning) to ensure a direct comparison against the primary baseline (\citet{lu2022multifidelity}). The experimental setup, including network architectures, largely mirrors that described in Section~\ref{sec:MF1}. However, to better leverage the more complex flow physics, the LF DeepONet is now trained for 800 epochs, while the subsequent HF fine-tuning phase remains at 400 epochs.

The results, presented in Table~\ref{tab:transfer_comparison_new}, again confirm the robustness of our proposed framework under these more demanding conditions. It is evident that the conventional approach by \citet{lu2022multifidelity} struggles to achieve a reasonable prediction error across all its configurations, indicating potential limitations when applied to such complex flow physics. In contrast, our proposed MF-DeepONet framework demonstrates much better performance, particularly when leveraging higher-quality LF data. Specifically, the four configurations using LF data at 64×64 and 128×128 resolutions with larger dataset sizes (200 and 300 samples) yield the most accurate and reasonable results. This highlights that the success of our framework is pronounced when the pre-training phase is supported by LF datasets that are both reasonably accurate in resolution and sufficient in quantity, enabling the model to effectively learn and transfer essential flow characteristics for high-fidelity prediction.

\begin{table}[htb!]
\centering
\setlength{\abovecaptionskip}{10pt} % Spacing
\renewcommand{\arraystretch}{1.}
\caption{Comparison of different MF-DeepONet architectures for new dataset. Single-fidelity results are shown for reference (e.g., ``Single-fidelity (32×32, 300)'' indicates 32×32 resolution with 300 samples).}

\label{tab:transfer_comparison_new}
\begin{tabular}{ccccc}
\hline
Architecture & LF Data Resolution & LF Data Size & Training Hour & MSE \\
\hline
Single-fidelity (32×32, 300) & --- & --- & 1.313 & 165.292 \\
Single-fidelity (64×64, 300) & --- & --- & 1.347 & 414.936 \\
Single-fidelity (128×128, 300) & --- & --- & 1.408 & 45.075 \\
Single-fidelity (256×256, 100) & --- & --- & 0.430 & 15.918 \\
\hline
\multirow{12}{*}{\makecell{\citet{lu2022multifidelity}\\(conventional MF-DeepONet)}}
& \multirow{4}{*}{32×32} & 50 & 0.498 & 50.228 \\
& & 100 & 0.550 & 60.736 \\
& & 200 & 0.770 & 50.059 \\
& & 300 & 0.976 & 55.436 \\
\cline{2-5}
& \multirow{4}{*}{64×64} & 50 & 0.447 & 51.301 \\
& & 100 & 0.564 & 52.415 \\
& & 200 & 0.831 & 50.653 \\
& & 300 & 1.019 & 49.644 \\
\cline{2-5}
& \multirow{4}{*}{128×128} & 50 & 0.469 & 50.367 \\
& & 100 & 0.605 & 48.326 \\
& & 200 & 0.809 & 39.131 \\
& & 300 & 1.054 & 32.995 \\
\hline
\multirow{12}{*}{\textbf{Proposed MF-DeepONet}}
& \multirow{4}{*}{32×32} & 50 & 0.598 & 29.997 \\
& & 100 & 0.686 & 25.242 \\
& & 200 & 0.915 & 23.959 \\
& & 300 & 1.117 & 20.610 \\
\cline{2-5}
& \multirow{4}{*}{64×64} & 50 & 0.576 & 30.048 \\
& & 100 & 0.715 & 23.012 \\
& & 200 & 0.885 & \textbf{13.919} \\
& & 300 & 1.058 & \textbf{12.028} \\
\cline{2-5}
& \multirow{4}{*}{128×128} & 50 & 0.586 & 23.745 \\
& & 100 & 0.708 & 20.957 \\
& & 200 & 0.914 & \textbf{18.302} \\
& & 300 & 1.111 & \textbf{12.364} \\
\hline
\end{tabular}
\end{table}

\subsection{Comparative Analysis of Physics-Guided Subsampling Strategy}
\label{sec:sampling_results_newdata}

We further investigate the efficacy of our proposed time-derivative guided subsampling strategy using these new challenging datasets, benchmarking it against the conventional residual-based sampling approach \cite{lu2021deepxde}. We apply these two subsampling methods to the four most accurate configurations identified in Table~\ref{tab:transfer_comparison_new} (highlighted in bold). As with our previous sampling experiments, we use a strategic sampling ratio of $r=0.1$.

The performance comparison of these sampling strategies is presented in Table~\ref{tab:dynamic_sampling_new}. While both subsampling methods offer comparable performance when using the 64x64 LF data, our physics-guided time-derivative strategy demonstrates a clear advantage when a higher-quality LF model is employed. Notably, for the case using 128x128 LF data with 300 samples, time-derivative subsampling yields a 14.20\% error reduction, which is more effective than the 11.63\% reduction from the residual-based method, while both have similar computational overhead. When the guiding LF model is of moderate quality (e.g., 64x64), its predictions contain more pronounced inaccuracies, making the error-driven residual sampling an effective corrective strategy. However, as the LF model becomes more accurate (e.g., 128x128), its ability to reliably identify physically dynamic regions becomes a more powerful form of guidance. In this regime, focusing on these areas of high temporal activity proves to be a more effective strategy for capturing complex dynamics than simply targeting the largest remaining (and now smaller) errors.

\begin{table}[htb!]
\centering
\setlength{\abovecaptionskip}{10pt} % Spacing
\renewcommand{\arraystretch}{1.1} % Reverted to 1.1
\caption{Performance comparison of different sampling strategies on the new datasets. "Before" refers to training with uniform subsampling.}
\label{tab:dynamic_sampling_new}
\begin{tabular}{ccccc}
\hline
Metric & Case & Before & \textbf{Time-Deriv. Sampling} & Residual Sampling \\
\hline
\multirow{4}{*}{MSE} & LF 64x64 w/ 200 samples & 13.919 & 14.484 (+4.06\%) & 14.482 (+4.04\%) \\
\cline{2-5}
& LF 64x64 w/ 300 samples & 12.028 & 11.929 (-0.82\%) & 11.889 (-1.16\%) \\
\cline{2-5}
& LF 128x128 w/ 200 samples & 18.302 & 18.147 (\textbf{-0.85\%}) & 18.386 (+0.46\%) \\
\cline{2-5}
& LF 128x128 w/ 300 samples & 12.364 & 10.609 (\textbf{-14.20\%}) & 10.926 (-11.63\%) \\
\hline
\multirow{4}{*}{Training hour} & LF 64x64 w/ 200 samples & 0.885 & 1.063 (+20.11\%) & 1.069 (+20.79\%) \\
\cline{2-5}
& LF 64x64 w/ 300 samples & 1.058 & 1.279 (+20.89\%) & 1.276 (+20.60\%) \\
\cline{2-5}
& LF 128x128 w/ 200 samples & 0.914 & 1.097 (+19.91\%) & 1.107 (+21.11\%) \\
\cline{2-5}
& LF 128x128 w/ 300 samples & 1.111 & 1.331 (+19.80\%) & 1.337 (+20.34\%) \\
\hline
\end{tabular}
\end{table}

\section{Conclusion}
\label{sec:conclusion}
This study presents a novel multi-fidelity DeepONet (MF-DeepONet) framework for efficient spatio-temporal flow field prediction with reduced high-fidelity data requirements. Through strategic architectural enhancements and physics-guided subsampling, we address critical challenges in operator learning for fluid dynamics applications. Our merge network architecture enables more complex feature interactions between operator and spatio-temporal coordinate spaces, achieving a 50.4\% reduction in prediction error compared to traditional dot-product approaches. The implementation of temporal positional encoding and efficient point-based sampling strategies further improves prediction accuracy by 7.57\% while dramatically reducing training time by 96\%. The transfer learning-based multi-fidelity framework leverages knowledge from pre-trained low-fidelity models to guide high-fidelity predictions, demonstrating a 43.7\% improvement in accuracy compared to single-fidelity training. Most significantly, our time-derivative guided subsampling strategy demonstrates two key advantages: (1) it reduces prediction error by 20.7\% compared to conventional uniform sampling when using the same amount of high-fidelity data, and (2) it achieves prediction accuracy comparable to a model trained on the complete set of available high-fidelity data, while using only 60\% of those training data---a crucial advancement for computationally expensive simulations. The robustness of these contributions is further validated on a second, more complex dataset with higher spatial resolutions and more intricate flow physics, where our proposed MF-DeepONet framework and time-derivative subsampling strategy continued to demonstrate superior or competitive performance against conventional benchmarks.

Despite these promising results, our framework also has limitations that warrant further investigation. First, the current transfer learning approach---which completely freezes branch and trunk networks---may be overly restrictive; more flexible layer-freezing strategies based on data similarity metrics could potentially yield further improvements. Additionally, while the time-derivative guided subsampling strategy proved effective for our flow problems, it requires broader validation across diverse physical systems with different temporal characteristics. Furthermore, its reliance on automatic differentiation introduces computational overhead, presenting opportunities for developing more efficient sampling implementations. Lastly, for broader industrial applicability, extending the framework to accommodate irregular domains, complex geometries, and moving boundaries represents an important direction for future work.

Future research will focus on several key areas to build upon the findings of this study. First, to improve the multi-fidelity framework itself, an automated layer-wise similarity analysis between low-fidelity and high-fidelity features could enable more nuanced transfer learning approaches with selective network component freezing. The observed training instability in some cases—attributable to the initial training of LF-DeepONet without guidance from HF data—highlights the need for refining the training strategy, which could serve as a promising direction for future research. Another important extension would be to investigate the robustness of MF-DeepONet when the LF and HF datasets are less strongly correlated, such as cases involving different Reynolds numbers or slight geometric domain changes. This would clarify the level of physical similarity required for effective transfer and broaden the framework's applicability in real-world scenarios. Furthermore, to enhance predictive robustness for time-evolving problems, future work will investigate the incorporation of physics-informed loss functions that more explicitly enforce the underlying Navier-Stokes dynamics during training. Second, to advance the sampling strategy, research can focus on developing physics-guided subsampling criteria beyond temporal derivatives---such as incorporating geometric information---to improve cross-domain generalizability and enable effective point selection in complex domains. The computational efficiency of these methods could also be enhanced through surrogate-based derivative estimation or alternative physics-guided metrics that avoid expensive automatic differentiation calculations. These advancements would further establish MF-DeepONet as a practical solution for real-time flow prediction in data-constrained environments.

\section*{Acknowledgments}
This work was supported by the Ministry of Science and ICT of Korea grant (No. RS-2024-00355857), and the National Research Council of Science \& Technology (NST) grant by the Korea government (MSIT) (No. GTL24031-000)

\section*{Data Availability Statement}

The data that support the findings of this study are available from the corresponding author upon reasonable request.

\appendix

\section{Automatic Mixed Precision (AMP)}\label{app:AMP}

AMP is a method that blends half-precision and full-precision computations to accelerate training and reduce memory usage, while maintaining stability. In simpler terms, it uses smaller data types (for example, 16-bit floats) during most calculations for speed and efficiency, then switches to larger data types (like 32-bit floats) for operations that risk losing too much detail. This strategy lowers the amount of memory required and speeds up math operations on modern hardware.

A crucial part of AMP is a safeguard mechanism that dynamically scales gradients to avoid floating-point overflow or underflow. In practice, a scaling factor is applied to gradients during backpropagation. After the updates are computed, this factor is reversed before adjusting the model’s parameters. If the training process detects an overflow---meaning the gradients became too large---the scaling factor is automatically reduced for the next round of updates, preserving numerical stability. Additional insights into mixed precision and gradient scaling can be found through sources such as the NVIDIA Developer Blog \cite{pytorch_amp}.

%Bibliography
% \bibliographystyle{unsrt}  
\bibliographystyle{plainnat}
\bibliography{references}

\end{document}